\documentclass[12pt,a4paper]{article}
\usepackage{jheppub}

\usepackage{dsfont}
\usepackage{amsmath,amssymb,amscd,amsfonts,mathtools}

\DeclareMathOperator{\Sech}{Sech}

\DeclareMathOperator{\Tanh}{Tanh}

\usepackage[toc,page]{appendix}
\usepackage{epsfig}
\usepackage{epstopdf}
\usepackage{latexsym}
\usepackage{graphicx}
\usepackage{subfigure}
\usepackage{booktabs}
\usepackage{bbm}
\usepackage{color}
\usepackage{physics}
\usepackage{tensor}
\usepackage{tikz}
\usetikzlibrary{matrix}
\usetikzlibrary{decorations.markings,calc,shapes,decorations.pathmorphing,patterns}
\usetikzlibrary{positioning}
\makeatletter
\def\@fpheader{\relax}
\makeatother

\subheader{\begin{flushright}
UTTG-23-2020\\
\end{flushright}}

\title{Warped Information and Entanglement Islands in AdS/WCFT}
\author{Elena Caceres$^{a}$, Arnab Kundu$^{b}$, Ayan K.~Patra$^{b}$, Sanjit Shashi$^{a}$}
\affiliation{$^a$Theory Group, Department of Physics, University of Texas, Austin, TX 78712, USA.}
\affiliation{$^b$Theory Division, Saha Institute of Nuclear Physics, HBNI, 1/AF Bidhannagar, Kolkata 700064, India.}

\emailAdd{elenac[at]utexas.edu, arnab.kundu[at]saha.ac.in, ayan.patra[at]saha.ac.in, sshashi[at]utexas.edu}

\abstract{We use the notion of double holography to study Hawking radiation emitted by the eternal BTZ black hole in equilibrium with a thermal bath, but in the form of warped CFT$_2$ degrees of freedom. In agreement with the literature, we find entanglement islands and a phase transition in the entanglement surface, but our results differ significantly from work in AdS/CFT in three major ways: (1) the late-time entropy decreases in time, (2) island degrees of freedom exist at all times, not just at late times, with the phase transition changing whether or not these degrees of freedom include the black hole interior, and (3) the physics involves a field-theoretic IR divergence emerging when the boundary interval is too big relative to the black hole's inverse temperature. This behavior in the entropy appears to be consistent with the non-unitarity of holographic warped CFT$_2$ and demonstrates that the islands are not a phenomenon restricted to black hole information in unitary setups.}

\begin{document}	
\maketitle
\flushbottom
	
\section{Introduction}\label{intro}

Recently, there has been a plethora of progress on the \textit{black hole information paradox}, which itself has been a central problem of quantum gravity for decades. At the semiclassical level, black holes can be said to evaporate through the pair production and emission of (thermal) Hawking radiation\cite{Hawking:1974sw}. Subsequent work by Hawking\cite{Hawking:1976ra} had found this process to be non-unitary---the final state of the gravitational system appears mixed, rather than pure, with the entanglement entropy monotonically increasing throughout evaporation. However, holographic considerations in the context of AdS/CFT had indicated that black hole evaporation should be unitary when phrased in the language of a dual field theory. Unitarity in this sense led to the \textit{Page curve}\cite{Page:1993wv,Page:2013dx} describing the entanglement entropy of the Hawking radiation over time; the Page curve for evaporation is characterized by rising, reaching a maximum at the so-called \textit{Page time}, then falling to zero by the \textit{evaporation time}. The Page time thus describes a phase transition. Essentially, Hawking's work yielded an ever-increasing entropy curve with a discontinuity at the evaporation time, instead, so, in the gravitational picture, the entangled degrees of freedom corresponding to Hawking radiation appear to vanish discontinuously.

Recent work in simple holographic models, reviewed in \cite{Almheiri:2020cfm}, is at odds with Hawking's past work. The most basic setup is presented by \cite{Almheiri:2019hni}. There, the full configuration is 2-dimensional and \textit{doubly holographic}. More specifically, the black hole is a solution to a dynamical gravitational theory in two dimensions called \textit{Jackiw-Teitelboim (JT) gravity}, and the matter fields are described by a holographic CFT$_2$. Furthermore, the black hole geometry is coupled via transparent boundary conditions to a zero-temperature \textit{bath}, \textit{i.e.} a CFT$_2$ on Minkowski space. This allows for matter fields to propagate away from the black hole, providing a mechanism for evaporation.

Such a setup is called doubly holographic because it lives in a theory with three equivalent descriptions, established by multi-sided application of holographic duality:\footnote{Such theories had also been discussed long ago in \cite{Karch:2000ct,Karch:2000gx}.}
\begin{itemize}
\item[(1)] a 2-dimensional, non-gravitating, zero-temperature \textit{boundary CFT (BCFT$_2$)} containing a 1-dimensional defect on which there exists a quantum mechanical dual to JT gravity (possibly $\mathcal{N} = 2$ SYK\cite{Forste:2017apw}),

\item[(2)] a 2-dimensional JT gravitational theory coupled to a CFT$_2$ and glued to the BCFT$_2$ of (1) at the latter theory's defect (turning it into an interface), with CFT degrees of freedom free to propagate through the interface due to transparent boundary conditions,

\item[(3)] a 3-dimensional, purely gravitational theory on AdS$_3$, but with a dynamical 2-dimensional boundary (identified as a \textit{Planck brane}\cite{Almheiri:2019hni}) containing the JT gravity degrees of freedom.
\end{itemize}

The transparent boundary conditions are pivotal for evaporation to take place. They are considered to be ``turned on" at some time $t = 0$. More specifically, \cite{Almheiri:2019hni} mentions activating them over some finite interval of time $\Delta t$. In (3), the reflecting boundary conditions for $t < 0$ translate to having a so-called \textit{Cardy brane} blocking the bath's bulk from the Planck brane's bulk, while turning on transparent boundary conditions smoothly is interepreted as having the Cardy brane move away from the conformal boundary and into the bulk.

To compute the Page curve in the semiclassical regime in description (2), one uses the machinery of \textit{quantum extremal surfaces (QES)}\cite{Engelhardt:2014gca}, the areas of which yield the fine-grained von Neumann entropy by minimizing the \textit{generalized entropy functional}. This functional, put simply, captures corrections to the classical entanglement entropy due to propagating bulk fields. When applying this machinery to an evaporating black hole in (2), one finds that the early-time QES is just the empty set. However, at some finite time, there is a phase transition at which the QES becomes a point disconnected from the conformal boundary\cite{Almheiri:2019psf}. This point is the boundary of a so-called \textit{entanglement island} which consists of degrees of freedom behind the horizon that are redundant with those of the emitted Hawking radiation. The need for this island in fully describing the Hawking degrees of freedom indicates that we essentially sacrifice locality in (2) to maintain unitarity in evaporation.

The work by \cite{Almheiri:2019hni} instead approaches the problem using description (3). Specifically, if the central charge of the matter CFT$_2$ is large, then the quantum part of the generalized entropy can be calculated to semiclassical order by application of the \textit{classical} RT\cite{Ryu:2006bv} or HRT\cite{Hubeny:2007xt} proposals in AdS$_3$/CFT$_2$ in the presence of a brane. There is a phase transition in this classical entanglement surface---the surface changes from one which entirely penetrates the bulk to another that intersects the Planck brane. Accounting for this phase transition produces a Page curve for the entropy, and the point on which the late-time classical entanglement surface intersects with the Planck brane is recognized as the boundary of the island.

The minimization of the generalized entropy functional may also be phrased in terms of the \textit{island rule}\cite{Almheiri:2019hni}, a modification to the classical RT/HRT prescriptions which includes the consideration of islands as part of the boundary region so as to ``complete" the homology constraint in a braneworld.
\begin{equation}
S(\mathcal{R}) = \text{\normalsize{min}}\left[\substack{\text{\normalsize{ext}}\\\mathcal{I}} \left(\frac{\text{Area}(\partial \mathcal{I})}{4G^{(2)}} + S_{\text{Bulk-2d}}(\mathcal{R} \cup \mathcal{I})\right)\right],\label{islandRule}
\end{equation}

Although the above doubly holographic setup is rather specific, the notion of entanglement islands is believed to be applicable to more general circumstances. For instance, one may consider non-evaporating setups, such a black holes \textit{in equilibrium} with an external, thermal bath, and simply look at the time evolution of entanglement surfaces.

Hartman and Maldacena considered surfaces in maximally-extended, eternal AdS black hole geometries\cite{Hartman:2013qma} that were meant to capture the entanglement between regions on each disjoint part of the overall conformal boundary. In other words, they computed the entanglement of degrees of freedom on the left part of the thermofield double with similar degrees of freedom on the right part. They describe these surfaces (relative to the $t = 0$ surface) as crossing the Einstein-Rosen bridge, thus increasing in area with time due to the growth of the bridge.

For an AdS$_d$ black hole in equilibrium with a bath and with holographic CFT$_d$ matter, one can use these \textit{Hartman-Maldacena surfaces} in the higher-dimensional theory to compute the entanglement entropy of Hawking radiation emitted through the CFT$_d$. The corresponding entropy curve, from $t = 0$, would thus appear to increase monotonically. This is an alternate version of the information paradox, since eternal black holes only contain a finite amount of information, thus defining an upper limit for how much entropy may be emitted.

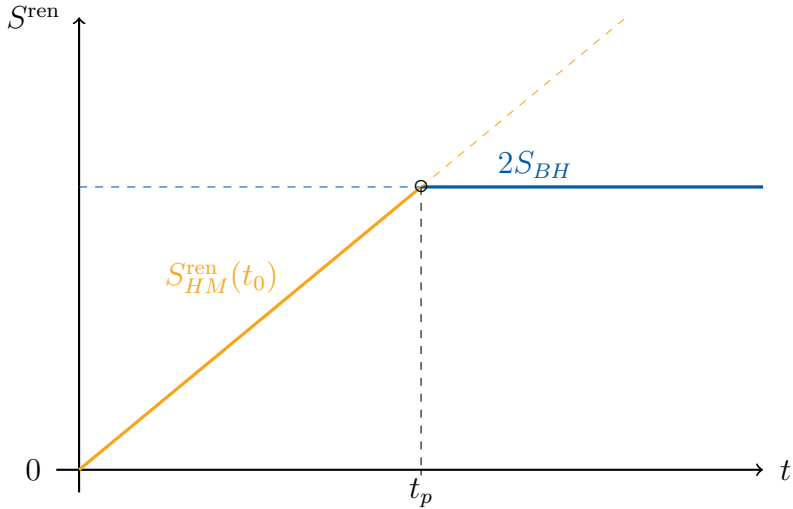
\begin{figure}
\centering
\begin{tikzpicture}[scale=1.5]
\draw[->,thick] (0,-0.2) to (0,4);
\draw[->,thick] (-0.2,0) to (6,0);

\node at (-0.4,4) {$S^{\text{ren}}$};
\node at (6.2,0) {$t$};

\draw[-,blue!30!teal,very thick] (3,2.5) to (6,2.5);
\draw[-,dashed,blue!30!teal] (0,2.5) to (3,2.5);

\draw[-,dashed,yellow!20!orange] (3,2.5) to (4.8,4);
\draw[-,yellow!20!orange,very thick] (0,0) to (3,2.5);
\node at (3,2.5) {$\circ$};
\draw[-,dashed] (3,2.5) to (3,0);
\draw[-] (3,0) to (3,-0.05);
\node at (3,-0.2) {$t_p$};

\node[yellow!20!orange] at (1.25,1.7) {$S_{HM}^{\text{ren}}(t_0)$};
\node[blue!30!teal] at (4,2.7) {$2S_{BH}$};

\node at (-0.4,0) {$0$};
\end{tikzpicture}
\caption{A sketch of the renormalized entanglement entropy of Hawking radiation emitted from an eternal AdS$_d$ black hole into a bath, as a function of time $t$. The orange curve $S_{HM}^{\text{ren}}(t)$ is the entropy of the Hartman-Maldacena surface, while the blue curve represents the maximum amount of information that can be emitted. The phase transition must occur by $t = t_p$, which is larger for more entropic black holes. This Page curve is only for $t \geq 0$, but we can also extend it to negative times. By time-reversal symmetry, the Page curve is actually symmetric about $t = 0$.}
\label{figs:eternalEntropy}
\end{figure}

This issue is resolved in \cite{Almheiri:2019yqk}, which introduces a second candidate surface in the higher-dimensional theory that produces islands on the AdS$_d$ black hole. By applying the standard minimality condition, one sees that this island-producing surface represents the late-time entropy, which has an upper bound of twice the Bekenstein-Hawking entropy.\footnote{Although the resulting plot of entropy versus time is not quite the Page curve obtained in the evaporating case, we still call this eternal variant a Page curve, as well, since it manifests from unitarity. The key point is that the entropy does not increase forever.} The Page curve of \cite{Almheiri:2019yqk} is shown Figure \ref{figs:eternalEntropy}. Furthermore, the island itself technically starts \textit{outside} of the horizon but still spans the interior.

This story is also discussed in \cite{Geng:2020qvw}, which considers the exterior patch of one side of a planar AdS$_5$ black hole. In just this patch, the Hartman-Maldacena surface is simply a radial line (suppressing the time and all but one of the non-radial spatial coordinates) which dives straight into the horizon. However, \cite{Geng:2020qvw} asserts the presence of a brane that allows for an additional candidate surface which does not penetrate the black hole horizon and, consequently, does not exhibit an increasing area in time in the maximally-extended geometry. Based on where this surface hits the brane, one observes the expected island from \cite{Almheiri:2019yqk} which starts outside of the horizon.

Another natural avenue to explore is higher-dimensional configurations\cite{Almheiri:2019psy}.\footnote{In this work, the authors consider $(3+1)$-dimensional \textit{Randall-Sundrum (RS) branes}\cite{Randall:1999vf} instead of Planck branes. Note that the exact form of the brane theory depends on the dimensionality, but the braneworld ideas of \cite{Randall:1999vf} form the basis of doubly holographic configurations.} In braneworld scenarios, one may also modify the brane tension\cite{Geng:2020qvw,Krishnan:2020fer}. However, if the brane theory is more than 3-dimensional, then one must consider the presence of massive gravitons on the brane\cite{Geng:2020qvw}. Other possible situations include geometries which may not have black holes at all by modifying the entropy functional\cite{Chen:2020uac,Chen:2020hmv}, islands in either asymptotically flat\cite{Gautason:2020tmk,Hashimoto:2020cas,Hartman:2020swn,Krishnan:2020oun} or de Sitter\cite{Balasubramanian:2020xqf,Sybesma:2020fxg} backgrounds, a bath at finite temperature\cite{Chen:2020jvn}, or a gravitating bath region\cite{Laddha:2020kvp,Geng:2020fxl}.

Even the doubly holographic braneworld setup is not essential in this story. Work has been done on realizing the appropriate unitary Page curve for a black hole by using the replica trick with the gravitational path integral. In doing so, one must consider contributions to the path integral by \textit{replica wormhole} geometries\cite{Penington:2019kki,Almheiri:2019qdq}, on top of the usual $n$-sheeted covering space. Along this vein, the idea of baby universes emerging behind the horizon has been used to describe the redundant island degrees of freedom in the interior of an evaporating black hole\cite{Akers:2019nfi,Balasubramanian:2020hfs}.

In this work, we follow yet another route; we change the theory describing the matter from a CFT$_2$ to something else. However, to keep the doubly holographic setup, the matter itself must be holographic, as well. Below, we elaborate on our motivation behind doing so.

In particular, we consider holographic \textit{warped CFT$_2$ (WCFT$_2$)}\cite{Hofman:2011zj,Detournay:2012pc}. The corresponding bulk is described either by AdS$_3$ or \textit{warped AdS$_3$ (WAdS$_3$)}\cite{Bengtsson:2005zj,Anninos:2008fx,Jugeau:2010nq}, as per the conjectured AdS$_3$/WCFT$_2$ and WAdS$_3$/WCFT$_2$ dualities. However, a generic WCFT$_2$ is non-local, and any holographic WCFT$_2$ is \textit{also} non-unitary. The details are discussed in Section \ref{wcftReview}, but concrete examples of WCFT$_2$ have been and continue to be explored in the literature. Specifically, \cite{Compere:2013aya} discusses \textit{chiral Liouville gravity}, a version of Liouville gravity in so-called chiral gauge, and suggests the existence of a (bosonic) quantum theory. Additionally, \cite{Hofman:2014loa} describes warped versions of free Weyl fermions\footnote{These are further explored by \cite{Castro:2015uaa}.} and $bc$ systems, and \cite{Jensen:2017tnb} discusses non-unitary scalar theories. Furthermore, \cite{Chaturvedi:2018uov} relates the complex SYK model to WCFT$_2$, noting that the former in the IR regime has the symmetry structure of the latter. Regardless, exploring the case of matter described by a WCFT$_2$ is a novel direction in the understanding of entanglement islands and their possible connections with \textit{unitarity} (or lack thereof) as well as \textit{non-locality} of the system. The prescriptions for holographic entanglement entropy in this arena have been explored in the literature\cite{Anninos:2013nja,Castro:2015csg,Song:2016gtd,Wen:2018mev,Gao:2019vcc,Apolo:2020bld,Apolo:2020qjm}.

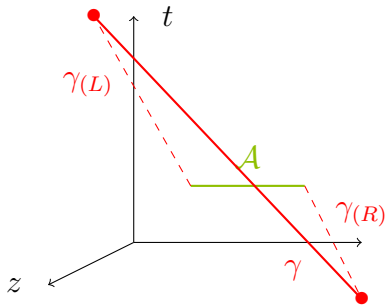
\begin{figure}
\centering
\begin{tikzpicture}[scale=1.5]
\draw[->] (0,0) to (0,2);
\draw[->] (0,0) to (-0.75,-0.375);
\draw[->] (0,0) to (2,0);

\node at (0.3,2) {$t$};
\node at (-1.05,-0.375) {$z$};

\draw[-, thick,color=black!25!lime] (0.5,0.5) to (1.5,0.5);
\node[color=black!25!lime] at (1,0.75) {$\mathcal{A}$};

\draw[-,color=red,dashed] (0.5,0.5) to (-0.35,2);
\draw[-,color=red,dashed] (1.5,0.5) to (2,-0.5);

\draw[-,color=red,thick] (-0.35,2) to (2,-0.5);

\node[color=red] at (-0.35,2) {$\bullet$};
\node[color=red] at (2,-0.5) {$\bullet$};

\node[red] at (-0.4,1.4) {$\gamma_{(L)}$};
\node[red] at (2,0.25) {$\gamma_{(R)}$};

\node[red] at (1.4,-0.25) {$\gamma$};
\end{tikzpicture}
\caption{A simplified sketch depicting a swing surface for a boundary interval $\mathcal{A}$ in a bulk space with time coordinate $t$ and radial coordinate $z$. The dashed lines $\gamma_{(L)}$ and $\gamma_{(R)}$ are the (null) ropes, while the solid line $\gamma$ is the (spacelike) bench.}
\label{figs:swing}
\end{figure}

Even more generally, it should be possible to utilize recent work on entanglement surfaces in the zoo of holographic dualities discussed by \cite{Wen:2018mev,Apolo:2020bld,Apolo:2020qjm}. They consider holographic dualities described by six assumptions listed in Section \ref{holoEnt}, all of which are satisfied by AdS/CFT, AdS$_3$/WCFT$_2$, and WAdS$_3$/WCFT$_2$. In such setups, the classical entanglement surface for a boundary region $\mathcal{A}$ is a so-called \textit{swing surface}. This is a union of null geodesics anchored to $\partial \mathcal{A}$, called \textit{ropes}, with a spacelike, codimension-$2$ surface in the bulk, called a \textit{bench}, connecting the ropes. The bench is selected via minimization, and the ropes allow for the entire swing surface to be homologous to $\mathcal{A}$. In Figure \ref{figs:swing}, we depict a sketch of such a surface.

In doubly holographic configurations where swing surfaces are used in place of the RT or HRT proposals, one can search for the emergence of islands by checking if the minimized swing surface for a boundary region in the bath intersects the brane---this is why we have chosen this particular prescription. Indeed, past prescriptions (particularly \cite{Song:2016gtd}) did not discuss the ropes and only treated WCFT$_2$ entanglement surfaces as a ``floating" geodesics (later identified as benches). Thus the role of homology only becomes clear when the ropes are involved, and we reiterate that the enforcement of homology in braneworlds is how islands are identified in the island rule \eqref{islandRule}.

Additionally, studying the behavior of entropy over time becomes a matter of understanding the behavior of swing surfaces corresponding to boundary intervals situated on different fixed-time slices. Consequently, we can use the swing surface prescription to study the time evolution of both entanglement islands and entanglement entropy.

To summarize, our immediate goal in this work is to perform an analysis of the entanglement surfaces in a doubly holographic setup akin to that of \cite{Almheiri:2019hni}, but with matter fields being described by a WCFT$_2$. Ultimately, using the prescription of \cite{Apolo:2020bld}, one finds that the eternal BTZ features entanglement islands at all times, with the accompanying entropy curve having a phase transition just as the Page curve for holographic CFT matter does. However, the curve itself is quite different from the Page curve, in that there is a constant piece \textit{followed by} a monotonically decreasing piece. We assert that this stems from the non-unitarity of holographic WCFT$_2$. In Section \ref{wcftReview}, we review WCFT$_2$ and discuss holography, concluding with a brief discussion of swing surfaces in AdS$_3$/WCFT$_2$. In Section \ref{braneworldWarped}, we introduce our doubly holographic model and compute both the islands and the entropy curve, deriving the time of the phase transition in the process. We also include an interpretation of our results, describing both how our results mesh with previous work and how this model, which includes non-unitary holographic matter, adds wrinkles to the general story about black hole information and islands.

\section{Warped CFT$_2$: Review}\label{wcftReview}

Here, we discuss the main differences between the symmetries of WCFT$_2$ and CFT$_2$. Additionally, we emphasize how WCFT$_2$ may be understood holographically, particularly in terms of an AdS$_3$ bulk. This duality is called \textit{AdS$_3$/WCFT$_2$}, and its dictionary includes a notion of holographic entanglement entropy similar to the RT and HRT proposals, termed the swing surface proposal by \cite{Apolo:2020bld,Apolo:2020qjm}.

\subsection{Symmetries, Unitarity, and Differences from CFT$_2$}\label{wcftSymmetry}

The seminal introduction to WCFT$_2$ is \cite{Detournay:2012pc}, which specifically explores symmetry structure, entropy, and holography. For now, we just review the global and local symmetry structure.

It is a well-known result of Polchinski that a CFT$_2$ on a Minkowski background can be obtained by imposing unitarity, Poincar\'e invariance, and global scale invariance\cite{Polchinski:1987dy}. Weakening our requirements a bit, it is natural to ask what happens when we have \textit{chiral scale invariance} in the right-moving sector alone; using light-cone coordinates $x^\pm = t \pm x$, combining such a symmetry with translational invariance means that we have the following three global symmetries (with $a,b \in \mathbb{R}$ and $\lambda > 0$):
\begin{equation}
x^+ \to x^+ + a,\ \ x^- \to x^- + b,\ \ x^- \to \lambda x^-.\label{chiralSym}
\end{equation}

Assuming \eqref{chiralSym} (which rules out Lorentz invariance and makes the theory \textit{non-relativistic}), \cite{Hofman:2011zj} finds that any unitary\footnote{Non-unitary WCFTs still feature chiral scale invariance. We will discard unitarity later.} theory whose dilatation operator has a discrete, non-negative spectrum must be either a CFT$_2$ or a WCFT$_2$. However, a CFT$_2$ will automatically have true scale invariance, so if we \textit{strictly} have chiral scale invariance, then the field theory is a WCFT$_2$.

That a WCFT$_2$ is characterized by strict chiral scale invariance has implications for the enhanced local symmetry structure of the quantum theory. Specifically, as we no longer have the full global conformal group, the local symmetries will not be described by two copies of the Virasoro algebra. Rather, a WCFT$_2$ is characterized by \textit{one} copy of the Virasoro algebra, corresponding to the chirally scale-invariant sector, and one copy of a U$(1)$ Kac-Moody algebra, corresponding to the purely translationally-invariant sector. \cite{Detournay:2012pc} writes the generators as $\{L_n,P_n\}_{n \in \mathbb{Z}}$, the central charge of the Virasoro algebra as $c$, and the level of the Kac-Moody algebra as $k$, so the canonical WCFT$_2$ algebra is,
\begin{align}
[L_n,L_m] &= (n-m)L_{n+m} + \frac{c}{12}n(n^2-1)\delta_{n+m},\\
[P_n,P_m] &= \frac{k}{2} n\delta_{n+m},\\
[L_n,P_m] &= -mP_{m+n}.
\end{align}

In a sense, the loss of Virasoro symmetry in the quantum theory relative to a true CFT$_2$ arises from the loss of full scale invariance.

Additionally, the symmetry structure appears to indicate that a generic WCFT$_2$ is \textit{non-local}. \cite{Song:2017czq} supports this by computing WCFT$_2$ correlation functions and finding that they ultimately fail to decay in the $U(1)$ direction.\footnote{\cite{Jensen:2017tnb} attempts to do away with this non-locality in the $U(1)$ direction, but in the process shows that the Hofman/Rollier WCFTs\cite{Hofman:2014loa} and their scalar counterparts admit infinite families of marginal non-local deformations. Thus, getting a local QFT requires an infinite number of tunings. Whether this QFT truly inherits the symmetry structure of the original WCFT$_2$ is unclear.} As this non-locality emerges by a discontinuous breaking of the full CFT$_2$ symmetry, there is no parameter which may smoothly interpolate between between WCFT$_2$ non-locality and CFT$_2$ locality.

Primary states are defined similarly to in a CFT$_2$. Specifically, take a state $\ket{p,h}$. Then it is said to be primary if, for $n > 0$,
\begin{equation}
P_n\ket{p,h} = L_n\ket{p,h} = 0,
\end{equation}
and,
\begin{equation}
P_0\ket{p,h} = p\ket{p,h},\ \ L_0\ket{p,h} = h\ket{p,h}.
\end{equation}

The descendant states are $P_{-n}\ket{p,h}$ and $L_{-n}\ket{p,h}$. Unitarity imposes constraints on the central charge, level, and weights ($p$ and $h$), however, and \cite{Detournay:2012pc} finds these constraints to be,
\begin{equation}
c \geq 1,\ \ k > 0,\ \ p \in \mathbb{R},\ \ h \geq \frac{p^2}{k}.\label{unitarityConst}
\end{equation}

\subsection{Holographic Duality}\label{wcftHolo}

The enhanced local symmetries of a WCFT$_2$ give us a hint about holographic duality. In AdS$_{3}$/CFT$_2$, the Brown-Henneaux boundary conditions\cite{Brown:1986nw} enhance the isometries of AdS$_{3}$, described by the (Minkowskian) \textit{$2$-dimensional global conformal group}, SO$(2,2)$, to two copies of the Virasoro algebra. Thus, if we wish to consider a gravity dual to WCFT$_2$, we must have a spacetime with boundary conditions under which the isometries enhance to the appropriate local symmetries.

There are two well-known proposed dualities meeting this criteria: \textit{AdS$_3$/WCFT$_2$} and \textit{WAdS$_3$/WCFT$_2$}. Both conjectured dualities are justified by asymptotic symmetry analysis, but in slightly different ways. In this work, we will rely on the former AdS$_3$/WCFT$_2$ duality. However, it is still worthwhile to understand how to arrive at WAdS$_3$/WCFT$_2$.

WAdS$_3$ is a non-maximally symmetric, Lorentzian spacetime of constant negative curvature. It is constructed by viewing AdS$_3$ as fiber bundle consisting of copies of $\mathbb{R}$ over AdS$_2$, then \textit{warping} those fibers by some positive factor. Details regarding the structure of WAdS$_3$ have been discussed in \cite{Bengtsson:2005zj,Anninos:2008fx,Jugeau:2010nq}. However, for our discussion, it is sufficient to note that the Lie algebra of Killing vectors loses some symmetry,
\begin{equation}
\text{so}(2,2) \xrightarrow{\text{warping}} \text{sl}(2,\mathbb{R}) \times \text{u}(1).\label{symmBreak}
\end{equation}

Thus, under the Dirichlet-Neumann boundary conditions discussed by \cite{Compere:2009zj}, \textit{all} of the spacetime symmetries of WAdS$_3$ are enhanced to some asymptotic symmetry algebra for a boundary field theory. Such an asymptotic algebra would be a Virasoro algebra plus a U$(1)$ Kac-Moody algebra, the hallmark of a WCFT$_2$. Summarily, the asymptotic symmetry algebra hints at WAdS$_3$/WCFT$_2$.

However, AdS$_3$/WCFT$_2$ is different, in that the relevant boundary conditions must \textit{not} preserve all of the spacetime symmetry. Such boundary conditions were found in \cite{Compere:2013bya} to be similar to the Dirichlet-Neumann boundary conditions characterizing WAdS$_3$/WCFT$_2$; in this context they are called \textit{Comp\`ere-Song-Strominger (CSS) boundary conditions}.

So, we have seen that a holographic dual for a WCFT$_2$ may be obtained in one of two ways. We either use a WAdS$_3$ bulk, or we change the boundary conditions of AdS$_3$ from those of Brown and Henneaux used for AdS$_3$/CFT$_2$. However, in the doubly holographic setup used to explore entanglement islands, it helps to use AdS$_3$, so as to not deviate too strongly from \cite{Almheiri:2019hni}. Regardless, note that prescriptions for holographic entanglement entropy in both AdS$_3$/WCFT$_2$ and WAdS$_3$/WCFT$_2$ have been explored in \cite{Anninos:2013nja,Castro:2015csg,Song:2016gtd,Wen:2018mev,Gao:2019vcc,Apolo:2020bld,Apolo:2020qjm}. For our purposes, however, we will primarily draw on the results of \cite{Apolo:2020bld,Apolo:2020qjm}, which build on \cite{Song:2016gtd}.

We conclude by noting that a holographic WCFT$_2$ with a dual AdS$_3$ bulk \textit{violates} unitarity, in the sense that there exist states with negative norm. This occurs because the Kac-Moody level $k$ is \textit{negative} for holographic theories\cite{Compere:2013bya,Apolo:2018eky}, directly violating \eqref{unitarityConst}. However, the central charge is still related to the bulk radius $L$ via the Brown-Henneaux equation,
\begin{equation}
c = \frac{3L}{2G^{(3)}} \gg 1.\label{bhCentralCharge}
\end{equation}

The authors of \cite{Apolo:2018eky} constrain holographic WCFT$_2$, despite the lack of true unitarity. They find that, despite the negative level, holographic WCFT$_2$ has primary states with either negative norm or imaginary Kac-Moody weight,
\begin{equation}
ip \in \mathbb{R}.\label{imagP}
\end{equation}

However, primaries for which \eqref{imagP} hold are mentioned in \cite{Apolo:2018eky} to have natural bulk interpretations as:
\begin{itemize}
\item[(1)] the AdS$_3$ vacuum,
\item[(2)] causal singularities in the form of closed timelike curves (CTCs),
\item[(3)] conical deficits and surpluses.
\end{itemize}

So, in short, although holographic WCFTs are not truly unitary, they are still under control so long as the primaries, at worst, satisfy \eqref{imagP} instead of having negative norm. It is thus sensible to discuss Hawking radiation via WCFT$_2$ degrees of freedom. As we are considering doubly holographic configurations, we will be assuming a non-unitary WCFT$_2$ with large central charge \eqref{bhCentralCharge} from here on out. Note that \cite{Alishahiha:2020qza} has observed both islands and a unitary Page curve for higher-derivative gravitational theories which may be non-unitary, but non-unitarity of the \textit{matter} sector appears to change the story in that we no longer have a Page curve.

\subsection{AdS$_3$/WCFT$_2$ and Holographic Entanglement Entropy}\label{holoEnt}

We now review the holographic prescription for classical entanglement entropy in AdS$_3$/WCFT$_2$, analogous to both the Ryu-Takayanagi and Hubeny-Rangamani-Takayanagi prescriptions of AdS/CFT. In particular, we focus on the application of the swing surface proposal discussed in \cite{Apolo:2020bld,Apolo:2020qjm} to the case of AdS$_3$/WCFT$_2$.\footnote{See also an earlier discussion of such entanglement surfaces conducted by \cite{Wen:2018mev}.} While swing surfaces can be used in other models, they provide a bridge between the ``floating" geodesics in prior literature on AdS$_3$/WCFT$_2$ holographic entanglement proposals\cite{Song:2016gtd} and the analogous (H)RT surfaces of AdS/CFT. Additionally, that the swing surface prescription incorporates a homology condition makes it rather convenient for finding islands, as we will see in Section \ref{braneworldWarped}.

The reader may worry about the validity of the notion of entanglement entropy in a system that is non-unitary. It is indeed not obvious that one can define a reduced density matrix in such cases. We do not offer a precise resolution of this issue. However, we emphasize that numerous recent works suggest that entanglement entropy may still hold for such systems. Towards that, we point the reader to \cite{Couvreur_2017} for a comprehensive study of a wide class of non-unitary CFTs, in which precise calculations of entanglement entropy can be performed. Furthermore, we only needed to rely on a geometric description, which is certainly well-defined from a bulk perspective. Therefore, we make a highly motivated assumption that the von Neumann entropy is well-defined for us.

We start with a rather general description of swing surfaces. \cite{Apolo:2020bld,Apolo:2020qjm} consider holographic models consisting of quantum gravity in $d+1$ dimensions and a $d$-dimensional field theory at the boundary, satisfying, among other possible consistency conditions (as in AdS/CFT\cite{ElShowk:2011ag}), the following criteria:
\begin{itemize}
\item[(1)] the bulk theory has a semiclassical description which may be written in terms of Einstein gravity,

\item[(2)] the field theory has a symmetry group $G$, and the vacuum state itself is invariant under some subgroup of $G$, generated by elements of the form $h_i$,

\item[(3)] there exist well-defined, consistent boundary conditions such that the bulk asymptotic symmetry group agrees with $G$,

\item[(4)] the bulk theory has a solution, identified as the dual to the vacuum state, whose Killing vectors $H_i$ correspond to the asymptotic generators $h_i$ at the boundary,

\item[(5)] the bulk and boundary theories' partition functions agree,

\item[(6)] ball-shaped regions of the boundary theory's vacuum configurations have a local modular Hamiltonian.
\end{itemize}

Conditions (1)-(5) appear rather standard, and in conjunction with condition (6) are used to generalize the Rindler method\cite{Casini:2011kv}, Lewkowycz-Maldecena\cite{Lewkowycz:2013nqa}, and Dong-Lewkowycz-Rangamani\cite{Dong:2016hjy} to so-called \textit{non-AdS holography}\footnote{This actually refers to holographic systems which depart from AdS/CFT specifically, so AdS$_3$/WCFT$_2$ would count.} in order to derive the swing surface proposal. The importance of the modular Hamiltonian in this procedure has been emphasized in earlier work by \cite{Wen:2018whg}.

The swing surface proposal starts as follows---for a boundary region $\mathcal{A}$, one first shoots null geodesics from each point $p \in \partial\mathcal{A}$, respectively labeled as $\gamma_{(p)}$. One does so by taking, near the boundary, the tangent vector of $\gamma_{(p)}$ to be an asymptotic Killing vector which reduces to a so-called \textit{approximate modular flow generator} at $p$, denoted by $\zeta^{(p)}$. This generator can be defined whenever there exist Rindler coordinates on a local neighborhood of $p$, in which case $\zeta^{(p)}$ is the generator of translations in local Rindler time. The trajectory of $\gamma_{(p)}$ away from the boundary is then defined by a corresponding approximate \textit{bulk} modular flow from $p$, ultimately terminating at a fixed point. These null geodesics are called \textit{ropes}.

Upon finding the ropes, one must subsequently compute the minimal, extremal, codimension-2, spacelike surface whose boundary is contained within the region bounded by the ropes. That is, when performing the extremization, we pick-out surfaces $X$ whose boundaries lie on $\bigcup_{p \in \partial\mathcal{A}} \gamma_{(p)}$, so they \textit{hang} from the ropes. Then we minimize over all such extremal surfaces. The resulting surface is called the \textit{bench} $\gamma$, and its area is the holographic entanglement entropy of $\mathcal{A}$.

Mathematically, denoting the entanglement entropy of $\mathcal{A}$ by $S_{\mathcal{A}}$,
\begin{equation}
S_{\mathcal{A}} = \min\left[\substack{\text{\normalsize{ext}}\\{X_{\mathcal{A}} \sim \mathcal{A}}} \frac{\text{Area}(X_{\mathcal{A}})}{4G^{(d+1)}}\right],\ \ X_{\mathcal{A}} = \left(\bigcup_{p \in \partial\mathcal{A}} \gamma_{(p)}\right) \cup X.
\end{equation}

$X$ denotes arbitrary codimension-2 surfaces which hang from the ropes, and $X_{\mathcal{A}} \sim \mathcal{A}$ means that the two surfaces are homologous, as is the case with (H)RT surfaces in AdS/CFT. The area ultimately reduces to that of the bench because each $\gamma_{(p)}$ is null, so only $\gamma$ has a nonzero contribution. See Figure \ref{figs:swing} for a simplified sketch of a swing surface. 

There is a subtlety with regards to IR regularization of the bulk radial coordinate in how different entanglement surfaces must be compared when exploring non-AdS holography. In AdS/CFT, the divergence structures of the areas of surfaces anchored to the boundary have been determined in \cite{Graham:1999pm}. Specifically in AdS$_3$, the length of a spacelike line exhibits a logarithmic divergence at the boundary. All lines that hit the boundary the same number of times thus have the same divergence structures, so a minimality constraint on surfaces homologous to $\mathcal{A}$ cannot be sensitive to a cutoff. As bulk IR divergences are interpreted via the holographic dictionary as boundary field-theoretic UV divergences, we may also say that the choice of entanglement surface is not affected by the field theory's UV cutoff.\footnote{Because of the UV/IR relations and to avoid ambiguity with another field-theoretic IR divergence which appears later in our work, we will refer to the cutoff for the radial coordinate as a (boundary) UV cutoff.}

However, swing surfaces are anchored to the boundary by the null ropes, so divergences in the area depend entirely on benches situated in the bulk. Such divergences must be treated carefully; when comparing different benches, one must also check how their UV divergence structures align.

As an aside, the swing surface proposal reduces to (H)RT in AdS/CFT. One finds that the fixed points of the bulk modular flow \textit{are} those of $\partial \mathcal{A}$ on the boundary\cite{Apolo:2020qjm}. As such, the ropes reduce to points, and the bench itself is homologous to the boundary as expected.

In AdS$_{3}$/WCFT$_2$, however, the ropes are non-trivial, and the bench manifests as a ``floating" geodesic, as described in \cite{Song:2016gtd}.\footnote{Because ``surfaces" are actually lines in AdS$_3$/WCFT$_2$, any ``area" is actually a length. We will employ this abuse of notation throughout our work.} We can see this explicitly by considering vacuum solutions in three dimensions. For such geometries, \cite{Apolo:2020bld} evaluates swing surfaces by working with the following metric (written in \textit{lightcone coordinates} $(\rho,u,v)$ and with the AdS$_3$ radius set to $1$),
\begin{equation}
ds^2 = \frac{d\rho^2}{4(\rho^2 - 4T_u^2 T_v^2)} + \rho du dv + T_u^2 du^2 + T_v^2 dv^2,\label{metLC}
\end{equation}
where $\rho^2 \geq 4T_u^2 T_v^2$ and $(u,v) \sim (u+2\pi,v+2\pi)$. The dual WCFT$_2$ lives on a cylinder at $\rho \to \infty$, but we can consider it to live on a plane, instead, if we decompactify $u$ and $v$.

The different vacuum solutions can be classified according to energy $\mathcal{E}$ and angular momentum $\mathcal{J}$,
\begin{equation}
\mathcal{E} = \frac{T_u^2 + T_v^2}{4G^{(3)}},\ \ \mathcal{J} = \frac{T_u^2 - T_v^2}{4G^{(3)}}.\label{energyMom}
\end{equation}

The specific types of backgrounds are,
\begin{align}
\text{AdS$_3$ vacuum:}\ &\mathcal{E} = -\frac{1}{8G^{(3)}},\ \ \mathcal{J} = 0,\\
\text{Conical defects:}\ &0 > \mathcal{E} > -\frac{1}{8G^{(3)}},\\
\text{BTZ black holes:}\ &\mathcal{E} \geq |\mathcal{J}| \geq 0.
\end{align}

To more concretely illuminate this discussion, we now explicitly write the swing surfaces in a static BTZ background for \textit{single} intervals. The expressions we obtain will be useful in our doubly holographic model, as well.

\subsubsection{Static BTZ Swing Surfaces}\label{staticBTZ}

For simplicity, we consider a static planar two-sided BTZ black hole, so that $T_u^2 = T_v^2$ (allowing us to take either to be the temperature) and $u$ and $v$ are decompactified. Then, denoting the black hole temperature as $T$, the horizon radius is,
\begin{equation}
\rho_h = 2T^2,\label{tempLC}
\end{equation}
and \eqref{metLC} (with $\rho \geq \rho_h$ and $u,v \in \mathbb{R}$) reads,
\begin{equation}
ds^2 = \frac{d\rho^2}{4(\rho^2 - \rho_h^2)} + \rho du dv + \frac{\rho_h}{2}(du^2 + dv^2).\label{btzMet}
\end{equation}

\begin{figure}
\centering
\begin{tikzpicture}[scale=1.5]
\draw[-,draw=none,fill=black!10] (1,1) to (1,-1) to (0,0) to (1,1); 
\draw[-,draw=none,fill=black!10] (-1,1) to (-1,-1) to (0,0) to (-1,1);

\draw[-,draw=none,fill=violet!10] (-1,1) to[bend right] (1,1) to (0,0) to (-1,1); 
\draw[-,draw=none,fill=violet!10] (1,-1) to[bend right] (-1,-1) to (0,0) to (1,-1); 

\draw[-,decoration = {zigzag,segment length = 1mm, amplitude = 0.25mm},decorate] (-1,1) to[bend right] (1,1);
\draw[-] (1,1) to (1,-1);
\draw[-,decoration = {zigzag,segment length = 1mm, amplitude = 0.25mm},decorate] (1,-1) to[bend right] (-1,-1);
\draw[-] (-1,-1) to (-1,1);

\draw[-,dashed,color=red] (-1,1) to (1,-1);
\draw[-,dashed,color=red] (1,1) to (-1,-1);

\draw[->] (-1.5,-0.5) to (-1.5,0.5);

\node at (-1.7,0) {$t$};

\draw[->,thick] (1.5,0) to (2.5,0);
\node at (2,-0.2) {\footnotesize$/\mathbb{Z}_2$};

\draw[-,draw=none,fill=black!10] (1+4,1) to (1+4,-1) to (0+4,0) to (1+4,1); 
\draw[-,draw=none,fill=black!10] (-1+4,1) to (-1+4,-1) to (0+4,0) to (-1+4,1);

\draw[-,draw=none,fill=violet!10] (-1+4,1) to[bend right] (1+4,1) to (0+4,0) to (-1+4,1); 
\draw[-,draw=none,fill=violet!10] (1+4,-1) to[bend right] (-1+4,-1) to (0+4,0) to (1+4,-1); 

\draw[-,decoration = {zigzag,segment length = 1mm, amplitude = 0.25mm},decorate] (-1+4,1) to[bend right] (1+4,1);
\draw[-] (1+4,1) to (1+4,-1);
\draw[-,decoration = {zigzag,segment length = 1mm, amplitude = 0.25mm},decorate] (1+4,-1) to[bend right] (-1+4,-1);
\draw[-] (-1+4,-1) to (-1+4,1);

\draw[-,dashed,color=red] (-1+4,1) to (1+4,-1);
\draw[-,dashed,color=red] (1+4,1) to (-1+4,-1);

\draw[-,draw=none,fill=white] (4,-1) to (4,1) to (5.7,1) to (5.7,-1) to (4,-1);
\end{tikzpicture}
\caption{The Penrose diagram for the maximally-extended, two-sided BTZ, with one of the spatial dimensions ($\phi$ in AdS-Schwarzschild \eqref{btzAS}) suppressed. Each gray patch is an exterior region, while the purple patches comprise the interior. This geometry exhibits a $\mathbb{Z}_2$ symmetry which proves useful in our calculations. Specifically, we only need to work within a single exterior region.}
\label{figs:penrose}
\end{figure}
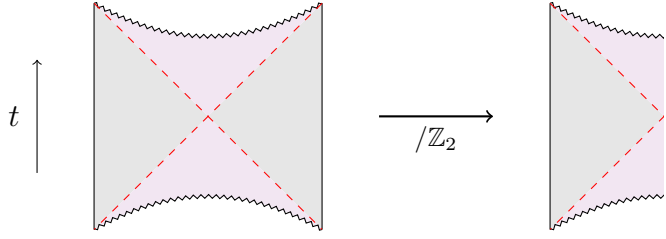

Recall that this would only cover a \textit{patch} of the maximally-extended geometry. Specifically, this metric describes one of the exterior regions in Figure \ref{figs:penrose}.

We consider a boundary interval $\mathcal{A}$ with endpoints,
\begin{equation}
\partial\mathcal{A} = \{(u_-,v_-),(u_+,v_+)\},\ \ l_u = u_+ - u_-,\ \ l_v = v_+ - v_-,\label{boundaryLC}
\end{equation}
so there exists two ropes, $\gamma^{\text{BTZ}}_{\pm}$, emanating from the points $(u_+,v_+)$ and $(u_-,v_-)$ respectively. In terms of a positive affine parameter $\lambda$, the ropes are,\footnote{Technically, we have rescaled the affine parameter relative to what is presented in \cite{Apolo:2020bld}, eliminating the constant momentum of the ropes.}
\begin{equation}
\gamma^{\text{BTZ}}_{\pm}(\lambda) = \begin{cases}
\rho(\lambda) = \sqrt{2\rho_h} \lambda + \rho_h,\vspace{0.2cm}\\
u(\lambda) = \mp\dfrac{1}{2}\sqrt{\dfrac{1}{2\rho_h}}\log\left(1 + \dfrac{\sqrt{2\rho_h}}{\lambda}\right) + u_{\pm} + O\left(\dfrac{1}{\rho_\infty^2}\right),\vspace{0.2cm}\\
v(\lambda) = \mp\dfrac{1}{2}\sqrt{\dfrac{1}{2\rho_h}}\log\left[\dfrac{2\rho_h \lambda(\lambda + \sqrt{2\rho_h})}{\rho_\infty^2}\right] + v_{\pm} + O\left(\dfrac{1}{\rho_\infty^2}\right).
\end{cases}\label{ropesLC}
\end{equation}

$\rho_\infty$ defines a UV cutoff in the boundary field theory. Taking $\lambda \to \infty$ corresponds to reaching the conformal boundary. However, there exists some finite, positive value for $\lambda > 0$ corresponding to reaching fixed points of the bulk modular flow, \textit{i.e.} the endpoints of the ropes.

In accordance with the former limit of $\lambda$, the affine parameter at the cutoff surface includes an $O(\rho_\infty)$ term,
\begin{equation}
\lambda_{\infty} = \frac{\rho_{\infty} - \rho_h}{\sqrt{2\rho_h}},
\end{equation}

Furthermore, the minimized bench from \cite{Apolo:2020bld} is a straight line,
\begin{equation}
\gamma^{\text{BTZ}} = \begin{cases}
\rho = \rho_h\coth\left(l_u\sqrt{\dfrac{\rho_h}{2}}\right),\vspace{0.2cm}\\
u = \dfrac{u_+ + u_-}{2},\vspace{0.2cm}\\
v \in \left[\dfrac{v_+ + v_- - \Delta v}{2},\dfrac{v_+ + v_- + \Delta v}{2}\right],
\end{cases}\label{benchLC}
\end{equation}
where $\Delta v$ is a cutoff-dependent, divergent term of the form,
\begin{equation}
\Delta v = l_v + \sqrt{\frac{2}{\rho_h}}\log\left[\frac{\rho_{\infty}}{\rho_h}\sinh\left(l_u \sqrt{\frac{\rho_h}{2}}\right)\right].
\end{equation}

As discussed in \cite{Apolo:2020qjm}, the bench hangs from the endpoints of the ropes fixed under the bulk modular flow. The affine parameter at the bench $\lambda_b$ is found by setting the radial coordinates of \eqref{ropesLC} and \eqref{benchLC} equal to one another,
\begin{equation}
\sqrt{2\rho_h}\lambda_b + \rho_h = \rho_h\coth\left(l_u\sqrt{\frac{\rho_h}{2}}\right) \implies \lambda_b = \sqrt{\frac{\rho_h}{2}}\left[\coth\left(l_u \sqrt{\frac{\rho_h}{2}}\right) - 1\right].\label{lambdaB}
\end{equation}

Additionally, equipped with the metric \eqref{btzMet}, we can compute the area of \eqref{benchLC} and, consequently, the associated holographic entanglement entropy as,
\begin{align}
\text{Area}(\gamma^{\text{BTZ}})
&= \sqrt{\frac{\rho_h}{2}}\Delta v = l_v \sqrt{\frac{\rho_h}{2}} + \log\left[\frac{\rho_\infty}{\rho_h}\sinh\left(l_u\sqrt{\frac{\rho_h}{2}}\right)\right],\label{areaBench}\\
S_{\mathcal{A}} &= \frac{\text{Area}(\gamma^{\text{BTZ}})}{4G^{(3)}}.\label{entanglementEntropy}
\end{align}

In order for this to line-up with prior results from WCFT$_2$ calculations, the cutoff $\rho_\infty$ is related to the WCFT$_2$ lattice spacing $\varepsilon$ by the matching,
\begin{equation}
\rho_\infty = \frac{\sqrt{2\rho_h}}{\varepsilon}.\label{cutoffUVIR}
\end{equation}

However, as our goal is to study black hole information dynamics, we need a notion of time. Thus, we introduce \textit{AdS-Schwarzschild coordinates} $(t,z,\phi)$.
\begin{align}
u &= \phi + t,\label{uPhit}\\
v &= \phi - t,\label{vPhit}\\
\rho &= \frac{1}{z^2} - \frac{1}{2z_h^2}.
\end{align}
where $z_h$ is the horizon coordinate, related to $\rho_h$ by,
\begin{equation}
\rho_h = \frac{1}{2z_h^2}.\label{horizonTrans}
\end{equation}

In AdS-Schwarzschild coordinates, \eqref{btzMet} is more recognizably one of the exterior patches ($z \leq z_h$) of a planar black hole described by a blackening factor $h(z)$,
\begin{equation}
ds^2 = \frac{1}{z^2}\left[-h(z) dt^2 + \frac{dz^2}{h(z)} + d\phi^2\right],\ \ h(z) = 1 - \frac{z^2}{z_h^2}.\label{btzAS}
\end{equation}

The $t$-coordinate ($t \in \mathbb{R}$) is time, and the $\phi$-coordinate ($\phi \in \mathbb{R}$) is a transverse spatial direction. In these coordinates, the temperature \eqref{tempLC} reads as,
\begin{equation}
T = \frac{1}{2z_h}.\label{tempAS}
\end{equation}

The conformal boundary is now at $z \to 0$. We will be concerned with surfaces analogous to those of the RT proposal, \textit{i.e.} entanglement surfaces for a boundary region at constant $t = t_0$. For such a constant-$t$ interval $\phi \in [\phi_-,\phi_+]$, \eqref{boundaryLC} becomes,
\begin{equation}
\partial\mathcal{A} = \{(t_0,\phi_-),(t_0,\phi_+)\},\ \ l_u = l_v = \phi_+ - \phi_-.\label{interval}
\end{equation}

The ropes become,
\begin{equation}
\gamma_{\pm}^{\text{BTZ}}(\lambda) = \begin{cases}
z(\lambda) = \dfrac{z_h}{\sqrt{1 + z_h \lambda}},\vspace{0.2cm}\\
\phi(\lambda) = \mp\dfrac{z_h}{2}\log\left(\dfrac{1 + z_h \lambda}{z_h^2 \rho_\infty}\right) + \phi_{\pm} + O\left(\dfrac{1}{\rho_\infty^2}\right),\vspace{0.2cm}\\
t(\lambda) = \pm\dfrac{z_h}{2}\log\left(\dfrac{\lambda}{z_h \rho_\infty}\right) + t_0 + O\left(\dfrac{1}{\rho_\infty^2}\right).
\end{cases}\label{ropesAS}
\end{equation}

Observe that, as $\lambda$ decreases from $\lambda_\infty$ to $\lambda_b$, the $\phi$-component of $\gamma_+^{\text{BTZ}}(\lambda)$ \textit{increases} montonically, while the $\phi$-component of $\gamma_-^{\text{BTZ}}(\lambda)$ \textit{decreases} monotonically.
\begin{equation}
-\pdv{\phi_{+}}{\lambda} = \frac{z_h^2}{2(1+z_h \lambda)} > 0,\ \ -\pdv{\phi_{-}}{\lambda} = -\frac{z_h^2}{2(1+z_h \lambda)} < 0.
\end{equation}

Meanwhile, as $\lambda$ decreases, the $t$-component of $\gamma_+^{\text{BTZ}}(\lambda)$ \textit{decreases} monotonically, while the $t$-component of $\gamma_-^{\text{BTZ}}(\lambda)$ \textit{increases} montonically.
\begin{equation}
-\pdv{t_{+}}{\lambda} = -\frac{z_h}{2\lambda} < 0,\ \ -\pdv{t_{-}}{\lambda} = \frac{z_h}{2\lambda} > 0.
\end{equation}

The bench can be found by transforming \eqref{benchLC}. However, as we are only concerned with its endpoints, we can also evaluate \eqref{ropesAS} at $\lambda_b$, which, in AdS-Schwarzschild coordinates, reads,
\begin{equation}
\lambda_b = \frac{1}{2z_h}\left[\coth\left(\frac{\phi_{+} - \phi_-}{2z_h}\right) - 1\right].
\end{equation}

The endpoints of the bench in AdS-Schwarzschild coordinates are thus,
\begin{equation}
\gamma_{\pm}^{\text{BTZ}}(\lambda_b) = \begin{cases}
z(\lambda_b) = z_h\sqrt{\dfrac{2}{1 + \coth\left(\frac{\phi_+ - \phi_-}{2z_h}\right)}},\vspace{0.2cm}\\
\phi(\lambda_b) = \phi_{\pm} \mp \dfrac{z_h}{2}\log\left[\dfrac{\coth\left(\frac{\phi_+ - \phi_-}{2z_h}\right) + 1}{2z_h^2 \rho_\infty}\right] + O\left(\dfrac{1}{\rho_\infty^2}\right),\vspace{0.2cm}\\
t(\lambda_b) = t_0 \pm \dfrac{z_h}{2}\log\left[\dfrac{\coth\left(\frac{\phi_+ - \phi_-}{2z_h}\right)-1}{2z_h^2 \rho_\infty}\right] + O\left(\dfrac{1}{\rho_\infty^2}\right).
\end{cases}\label{benchAS}
\end{equation}

As opposed to RT surfaces in AdS/CFT, the entire swing surface does not lie on a constant-$t$ surface. Even the bench reaches across different constant-$t$ slices. This makes sense since one of the ropes is directed towards future null infinity, while the other rope is going towards past null infinity.

Lastly, the area of $\gamma^{\text{BTZ}}$ should not change under a coordinate transformation, so we may simply plug \eqref{horizonTrans} and \eqref{interval} into \eqref{areaBench} to write,
\begin{equation}
\text{Area}(\gamma^{\text{BTZ}}) = \frac{\phi_{+}-\phi_-}{2z_h} + \log\left[2z_h^2 \sinh\left(\frac{\phi_+ - \phi_-}{2z_h}\right)\right] + \log\rho_\infty.
\end{equation}

\section{Warped Entanglement Entropy in a Braneworld}\label{braneworldWarped}

After our review of holographic entanglement entropy in AdS$_3$/WCFT$_2$, we are ready to discuss doubly holographic models. We start by coupling 2-dimensional dynamical dilatonic gravity (a ``JT-like" theory) to holographic WCFT$_2$ degrees of freedom which extend into a non-gravitating bath via eternal transparent boundary conditions. Furthermore, the bath is taken to be in thermal equilibrium with the gravitational region. As a result, no gravitational evaporation occurs, and the setup is eternal. That being said, there would still be emitted Hawking degrees of freedom in the form of WCFT$_2$ matter.

We will demonstrate that there are two candidate swing surfaces, both featuring islands, with a phase transition between them. The time of this phase transition, however, will be seen to depend on an IR cutoff in the field theory. Furthermore, we obtain an entropy curve which differs radically from the Page curve of \cite{Almheiri:2019yqk}, in that it monotonically \textit{decreases} after the phase transition. Despite some passing similarities to previous work in AdS/CFT, we interpret our model as demonstrating that islands reaching \textit{behind} a black hole horizon need not be sufficient to prevent an information paradox in the entropy curve (possibly due to \textit{non-unitarity} in the matter sector). However, islands are still seen to be needed to appropriately describe entanglement entropy at all, so the idea that Hawking radiation generally requires redundant degrees of freedom lives on. That being said, we will also find that there are \textit{always} islands, a strong departure from AdS/CFT which we attribute to the coupling of gravity to a \textit{non-local} field theory.

\subsection{Warped Doubly Holographic Setup and Hawking Radiation}\label{doubleHolo}

Taking inspiration from the doubly holographic model of \cite{Almheiri:2019hni}, we consider a similar model using AdS$_3$/WCFT$_2$. In particular, we consider a 2-dimensional theory of dilatonic gravity $I_{WJT}$ coupled to holographic WCFT$_2$ matter $I_{W}$,\footnote{We are technically considering the semiclassical regime of some quantum gravity theory, so the metric, as a field, will not have any quantum backreaction, whereas matter sector observables are treated as expectation values.}
\begin{equation}
I[g_{ij}^{(2)},\varphi,\chi] = I_{WJT}[g_{ij}^{(2)},\varphi] + I_{W}[g_{ij}^{(2)},\chi].\label{2dSystem}
\end{equation}

The matter fields are represented by $\chi$, the gravity dilaton by $\varphi$, and the background metric by $g_{ij}^{(2)}$. Unlike Einstein gravity, \eqref{2dSystem} ultimately describes a 2-dimensional, \textit{dynamical} theory of gravity containing matter, thanks to the introduction of the dilaton.

In Appendix \ref{JTgravity}, we discuss how the analogous theory using JT gravity and holographic CFT$_2$ is solved classically by a fixed AdS$_2$ background \eqref{eomBackground2}, with the dilaton being coupled to the CFT$_2$ stress tensor \eqref{stress2SC} and having a particular boundary condition \eqref{dilatonBC}. We \textit{assume} these core elements---that there exists a classical saddle point of \eqref{2dSystem} which admits an AdS$_2$ background and which features a dilaton (subject to the dilaton boundary condition in \eqref{dilatonBC}) coupled entirely to the WCFT$_2$ stress tensor as \eqref{stress2SC}. However, we make no further assumptions about the form of the action, so we simply say that the gravity theory is \textit{JT-like}.

Additionally, we couple the ``end" of the gravitational region to a non-gravitating, flat bath, but with transparent boundary conditions at the interface. This allows for the WCFT$_2$ degrees of freedom to propagate out as Hawking radiation.

Because the WCFT$_2$ is holographic, we can consider a higher-dimensional dual description. Likewise, under the holographic principle, we can treat the JT-like gravity degrees of freedom as living on one dimension, instead of two. This amounts to the following triality,
\begin{itemize}
\item[(1)] a boundary WCFT$_2$, on whose 1-dimensional boundary there exists a quantum mechanical holographic dual to the JT-like theory,

\item[(2)] the 2-dimensional gravitational region coupled to a matter WCFT$_2$ with some UV cutoff $\varepsilon$ and a non-gravitating bath region, with transparent boundary conditions imposed at the interface, 

\item[(3)] Einstein gravity on locally-AdS$_3$ background subject to CSS boundary conditions and with a dynamical, JT-like gravitating brane.
\end{itemize}

The third picture (3) is by far the most straightforward one and where we perform our computations. The saddles here in a sense justify our assumption about the saddle in (2) including an AdS$_2$ background, since the branes may be taken from an AdS$_2$ slicing of the AdS$_3$ bulk. In doing so however, note that the CSS boundary conditions would also have to convey to the brane.

One can realize a thermal state in (1) to be dual to an AdS$_3$ black hole in (3) with a brane present. More specifically, we may consider a \textit{probe brane}\footnote{While it is possible to consider other types of branes with tension\cite{Takayanagi:2011zk}, probe branes are easier to work with.} which does not backreact with the bulk, as in \cite{Geng:2020qvw}. Consequently, the induced metric on the gravitating brane will be that of an AdS$_2$ black hole with the same radius as the geometry of (3), so this configuration in (2) is precisely a black hole coupled to and in equilibrium with a thermal bath. Similarly, the vacuum state in (1) arises by taking empty AdS$_3$ in (3) instead, thus inducing an empty AdS$_2$ geometry on the brane.

Furthermore the nuances of coupling of WCFT$_2$ to curved backgrounds is not entirely clear\cite{Hofman:2014loa}, but it is the view of the 2-dimensional system as being induced by the more straightforward 3-dimensional picture that saves us. Our analysis is thus in the latter.

As in \cite{Almheiri:2019hni}, we would expect that the appropriate prescription to compute the entropy of the Hawking radiation in (2) under the semiclassical limit would be the machinery of generalized entropy and quantum extremal surfaces\cite{Engelhardt:2014gca}. However, it is simpler to consider (3), because, from this perspective, a QES is well-approximated by a classical surface allowed to end on the brane. For the model in \cite{Almheiri:2019hni} in which the matter is described by a holographic CFT$_2$, this is said to be because the number of degrees of freedom (described by $c$) is large. Consequently, in the following generalized entropy functional evaluated over 0-dimensional collections of points $y$ in the 2-dimensional bulk,\footnote{One considers the interval $\mathcal{A}$ to be some bath radiation region union intervals in the 2-dimensional bulk with boundary points in $y$. $y$ may also be the empty set.}
\begin{equation}
S_{\text{gen}}(y) = \sum_{p \in y} \frac{\varphi(p)}{4G^{(2)}} + S_{\text{Bulk-}2d}[\mathcal{A}],\label{genEnt}
\end{equation}
the classical term in $S_{\text{Bulk-}2d}[\mathcal{A}]$ dominates over higher-order quantum fluctuations, allowing it to be written as some area in the 3-dimensional bulk by holography. Meanwhile the extra $\varphi$ contribution comes from the area of the endpoints of any islands that appear, these endpoints being identified as the intersection between the brane and the entanglement surface.

This argument does not change for a holographic WCFT$_2$. We still take $c$ to be large, so it should have a large number of degrees of freedom. However, the key difference is that $S_{\text{Bulk-}2d}[\mathcal{A}]$ instead becomes well-approximated by the area of a swing surface in the 3-dimensional braneworld.

Our search for islands will be as follows: in the 3-dimensional bulk geometry of (3), we define a large, constant-$t$ \textit{radiation region} $\mathcal{R}$ on the conformal boundary. The entanglement entropy of Hawking radiation is computed by first applying the swing surface proposal to $\mathcal{R}$, modified to accommodate a probe brane. Because the probe brane does not backreact onto the geometry, this modified prescription amounts to finding the swing surface with no brane present, then excising the geometry on one side of the brane. We then consider whether the swing surface intersects with the brane itself. Based on the \textit{island rule} of \cite{Almheiri:2019hni},
\begin{equation}
S(\mathcal{R}) =  \text{\normalsize{min}}\left[\substack{\text{\normalsize{ext}}\\\mathcal{I}} \left(\frac{\text{Area}(\partial \mathcal{I})}{4G^{(2)}} + S_{\text{Bulk-2d}}(\mathcal{R} \cup \mathcal{I})\right)\right],
\end{equation}
the appropriate entanglement island is identified as the region $\mathcal{I}$ such that $\mathcal{R} \cup \mathcal{I}$ is homologous to the piece of the swing surface remaining after excision.

We will observe that, for the eternal planar BTZ, there are two candidate swing surfaces. Both surfaces feature islands, and there is a phase transition between them. However, the transition time depends on the size of $\mathcal{R}$ relative to the black hole's inverse temperature $\beta$: the larger $\mathcal{R}$ is, then the earlier the phase transition occurs. The early-time surface has islands strictly outside of the horizon, while the late-time surface has an island which runs across the black hole interior.

\subsection{Radiation Surface Candidates in the BTZ}\label{radSurfBTZ}

Following the lead of \cite{Geng:2020qvw}, we again go back to the static planar two-sided BTZ \eqref{btzAS}, but with a probe brane $\phi = 0$ present. As for the radiation region, we take $t = t_0$ and two copies of the interval $[\phi_r,\phi_{IR}]$ (one on each side of the black hole). Here, $\phi_{IR}$ is a large cutoff relative to $z_h$, while $\phi_r > 0$. We also define the difference,
\begin{equation}
l_{\phi} = \phi_{IR} - \phi_r \gg z_h = \frac{\beta}{2},
\end{equation}

For now, we keep $l_\phi/z_h$ finite. This is a regularization procedure; we wish to prevent a possible field-theoretic IR divergence associated with the size of the boundary interval. We denote the radiation region as $\mathcal{R}(t_0)$.

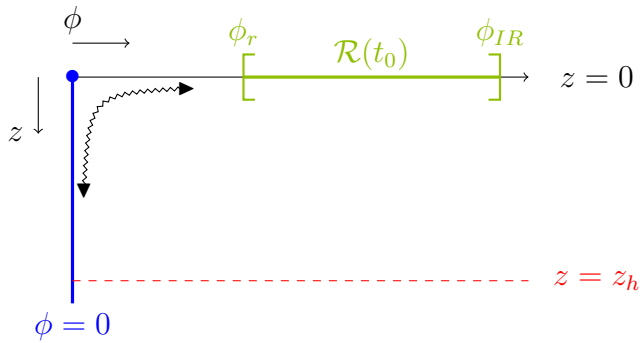
\begin{figure}
\centering
\begin{tikzpicture}[scale=1.5]
\draw[->] (0,2) to (2+2,2);
\draw[-,very thick,blue] (0,2) to (0,0);

\draw[->] (0,2.3) to (0.5,2.3);
\node at (0,2.5) {$\phi$};

\draw[->] (-0.3,2) to (-0.3,1.5);
\node at (-0.5,1.5) {$z$};

\node at (0,-0.2) {\textcolor{blue}{$\phi = 0$}};

\draw[red,-,dashed] (0,0.2) to (2+2,0.2);

\node at (2+2.6,0.2) {\textcolor{red}{$z = z_h$}};
\node at (2+2.6,2) {$z = 0$};

\draw[-,very thick,black!25!lime] (1.5,2) to (1.75+2,2);
\node at (2.625,2.2) {\textcolor{black!25!lime}{$\mathcal{R}(t_0)$}};

\draw[-,black!25!lime,thick] (1.6,2.2) to (1.5,2.2) to (1.5,1.8) to (1.6,1.8);
\node at (1.5,2.4) {\textcolor{black!25!lime}{$\phi_r$}};

\draw[-,thick,black!25!lime] (3.65,2.2) to (1.75+2,2.2) to (1.75+2,1.8) to (3.65,1.8);
\node at (1.75+2,2.4) {\textcolor{black!25!lime}{$\phi_{IR}$}};

\draw[->,decoration = {zigzag,segment length = 1mm, amplitude = 0.25mm},decorate] (0.1,1) .. controls (0+0.15,2-0.15) .. (1,1.9);
\draw[->] (0.1,1.01) to (0.1,1);
\node[rotate=180,scale=0.75] at (0.1,1) {$\blacktriangle$};
\node[rotate=-90,scale=0.75] at (1,1.9) {$\blacktriangle$};

\node[color=blue] at (0,2) {$\bullet$};
\end{tikzpicture}
\caption{The exterior of the planar BTZ at $t = t_0$ with the horizon at $z = z_h$ (in red) and a brane placed at $\phi = 0$ (in blue), on which the 3-dimensional geometry induces an AdS$_2$ black hole with the same horizon $z = z_h$. The Hawking radiation (depicted by a wavy arrow) is free to propagate between the AdS$_2$ black hole and the bath. We are interested in the swing surface corresponding to the radiation region (in green, but technically including another interval on the other side of the horizon).}
\label{figs:braneworld}
\end{figure}

One of the exterior regions for the above configuration is depicted in Figure \ref{figs:braneworld}. Additionally, the induced metric on $\phi = 0$ is an AdS$_2$ black hole with a horizon at $z = z_h$ and the radius still set to $1$,
\begin{equation}
ds^2|_{\phi = 0} = \frac{1}{z^2}\left[-h(z) dt^2 + \frac{dz^2}{h(z)}\right].
\end{equation}

By analyzing \eqref{ropesAS} (setting $\phi_+ = \phi_{IR}$ and $\phi_- = \phi_{r}$), we can see that the rope emanating from $\phi_{IR}$ will never cross the brane, since that rope's $\phi$-component monotonically increases. However, as the rope emanating from $\phi_r$ has a monotonically decreasing $\phi$-component, it may cross $\phi = 0$ precisely once at some allowed value of the affine parameter. We find this to be,
\begin{equation}
\lambda_0 = -\frac{1}{z_h} + z_h \rho_\infty e^{-2\phi_r/z_h} + O\left(\frac{1}{\rho_\infty}\right).\label{affineAt0}
\end{equation}

Observe that this intersection point is always near the cutoff surface. To see why, we find that the radial coordinate of both ropes at $\lambda = \lambda_0$ is,
\begin{equation}
z_\pm(\lambda_0) = \frac{1}{\sqrt{\rho_\infty}} \exp\left[\frac{\phi_r}{z_h} + O\left(\frac{1}{\rho_\infty^2}\right)\right] \xrightarrow{\rho_\infty \to \infty} 0.\label{radialRopes}
\end{equation}

We are ready to discuss the benches. Without a brane present, there are two possibilities---we may either connect ropes on \textit{opposite} sides of the horizon by two extremal, spacelike lines, or we may connect the ropes on the \textit{same} sides by using two copies of the single-interval bench discussed in Section \ref{staticBTZ}. We refer to the former as the \textit{Hartman-Maldacena swing}, in analogy to the Hartman-Maldacena surface of \cite{Hartman:2013qma} which increases in area with time and thus gives rise to an information paradox on its own. We refer to the swing surfaces in the latter case as the \textit{exterior swings}.

Upon introducing the brane, both candidate swing surfaces are truncated. This is because the ropes emanating from $\phi_r$ end in the $\phi < 0$ region. For the Hartman-Maldacena swing, the entire bench sitting in $\phi < 0$ is thus removed with much of the ropes. Meanwhile, portions of the exterior swings' benches and ropes (from $\phi_r$) hit the $\phi = 0$ brane, so they get cut-off, as well. As we will discuss in Section \ref{islandsBTZ}, this indicates that both surfaces have entanglement islands.

Because there are two candidates, we must consider both and take the one that gives a smaller area. We do this below.

\subsubsection{Hartman-Maldacena Swing}\label{hmSwing}

First, starting with the Hartman-Maldacena swing \textit{with} the brane present, the only spacelike contribution to the area is the line connecting the ropes emanating from $\phi_{IR}$. This line consists of two constant-$t$, constant-$\phi$ intervals from the ropes to the horizon (one in each exterior region) and a time-dependent piece in the interior.

Using \eqref{benchAS}, the metric, and the $\mathbb{Z}_2$-symmetry in Figure \ref{figs:penrose}, this area is,
\begin{equation}
A_{HM}(t_0)
= 2\int_{z(\lambda_b)}^{z_h} \frac{dz}{z\sqrt{h(z)}} + F(t_0)
= 2\Tanh^{-1}\left[e^{-l_\phi/(2z_h)}\right] + F(t_0),\label{HMarea1}
\end{equation}
where $F(t_0)$ is the interior contribution for $\mathcal{R}(t_0)$.

From \eqref{benchAS}, the exterior portion of this bench resides at the time (neglecting $O(1/\rho_\infty)$ terms),
\begin{equation}
t_+(\lambda_b) = t_0 + \frac{z_h}{2}\log\left[\frac{\coth\left(\frac{l_\phi}{2z_h}\right) - 1}{2z_h^2}\right] - \frac{z_h}{2}\log\rho_\infty.\label{timeHM}
\end{equation}

So, the bench is technically anchored to the slice at boundary time $t \to -\infty$ (approaching past infinity). However, by time-reversal symmetry of the two-sided BTZ geometry, this area is equal to that of a surface located at time,
\begin{equation}
\tilde{t}_+ = -t_+(\lambda_b) = -t_0 - \frac{z_h}{2}\log\left[\frac{\coth\left(\frac{l_\phi}{2z_h}\right) - 1}{2z_h^2}\right] + \frac{z_h}{2}\log\rho_\infty.
\end{equation}

Thus, from the area expression given in \cite{Hartman:2013qma} (worked out specifically for three dimensions in Appendix \ref{areaHM}), we have that,
\begin{equation}
F(t_0) = \frac{2 \tilde{t}_+}{z_h} = -\frac{2t_0}{z_h} - \log\left[\frac{\coth\left(\frac{l_\phi}{2z_h}\right) - 1}{2z_h^2}\right] + \log\rho_\infty.
\end{equation}

Expanding the IR-divergent terms about $l_\phi/z_h \to \infty$, we write \eqref{HMarea1} as,
\begin{equation}
A_{HM}(t_0) = -\frac{2t_0}{z_h} + 2\log z_h + \log\rho_\infty + \frac{l_\phi}{z_h} + O\left(\frac{z_h}{l_\phi}\right).\label{HMfinal}
\end{equation}

We observe that the Hartman-Maldacena bench has two divergences, with the logarithmic UV divergence stemming from its infinite area in the bulk and the linear IR divergence coming from the length of the boundary interval.

That we have a linearly decreasing area is because the bench sits near past infinity \eqref{timeHM} and gets ``dragged" forward in time as $t_0$ increases. Put differently, although the area is formally infinite due to the divergences, the bench is crossing the part of the Einstein-Rosen bridge which linearly \textit{decreases} in time.

\subsubsection{Exterior Swings}\label{extSwing}

The next area to compute is that of the two exterior swings which arise from connecting ropes in the same exterior regions prior to excision. It is easier to compute the area of just one of their benches in lightcone coordinates \eqref{metLC}, double the result, then convert to AdS-Schwarzschild, with,
\begin{eqnarray}
u_+ = \phi_{IR} + t_0,\ \ &u_- = \phi_r + t_0,\label{uPMRad}\\
v_+ = \phi_{IR} - t_0,\ \ &v_- = \phi_r - t_0.
\end{eqnarray}

The full bench in question is described in lightcone coordinates by \eqref{benchLC}. Observe that it is a constant-$\rho$, constant-$u$ surface. So, when we excise the $\phi < 0$ region, we only care about the value of $v$ at the $\phi = 0$ brane. This is found by noting,
\begin{equation}
\phi = \frac{u+v}{2} = 0 \implies u = -v,
\end{equation}
which indicates that, when this bench hits $\phi = 0$, the new endpoint remaining after the excision has the $v$-value,
\begin{equation}
v = -\frac{u_+ + u_-}{2}.
\end{equation}

Thus, the combined area of the two exterior surfaces is,
\begin{align}
A_{E}
&= 2\sqrt{\frac{\rho_h}{2}}\left(\frac{v_+ + v_- + \Delta v}{2} + \frac{u_+ + u_-}{2}\right)\nonumber\\
&= \frac{2\phi_r}{z_h} + \frac{3l_\phi}{2z_h} + \log\left[2z_h^2 \sinh\left(\frac{l_\phi}{2z_h}\right)\right] + \log\rho_\infty.\label{songArea1}
\end{align}

As in \eqref{HMfinal}, we expand the IR-divergent terms about $l_\phi/z_h \to \infty$ to rewrite \eqref{songArea1} as,
\begin{equation}
A_E = \frac{2\phi_r}{z_h} + 2\log z_h + \log\rho_\infty + \frac{2l_\phi}{z_h} + O\left(\frac{z_h}{l_\phi}\right).\label{songFinal}
\end{equation}

We again have two divergences: a logarithmic UV divergence in $\rho_\infty$ and a linear IR divergence in $l_\phi/z_h$.

\subsubsection{Area Difference between Swings}\label{compSwings}

We now compare the areas of the two swings (amounting to the areas of their benches) so as to find the minimal one. Note that the UV divergence, which emerges when $\rho_\infty \to \infty$, is of the same order in both \eqref{HMfinal} and \eqref{songFinal}. This also occurs in AdS/CFT and, consequently, the area \textit{difference} becomes a UV-finite quantity. The cutoff surface $\rho = \rho_\infty$ (and the corresponding field-theoretic lattice cutoff $\varepsilon > 0$) does not carry any true physical meaning in the choice of the entanglement surface.

However, the IR divergences, despite being linear in both areas, do \textit{not} cancel when we consider the difference. Specifically, we define,
\begin{align}
\Delta A(t_0)
&= A_{E} - A_{HM}(t_0)\nonumber\\
&= \frac{2}{z_h}(\phi_r + t_0) + \frac{l_\phi}{z_h} + O\left(\frac{z_h}{l_\phi}\right).\label{areaDiff}
\end{align}

Now, we compare the two candidate surfaces more thoroughly, both examining \eqref{areaDiff} and understanding their islands on the $\phi = 0$ brane.

\subsection{Entropy Curve and Islands}\label{islandsBTZ}

Equipped with the two candidate surfaces and their area difference, we discuss the entanglement entropy of $\mathcal{R}(t_0)$ as a function of $t_0$, with particular emphasis placed on the phase transition between the two surfaces. As usual, the appropriate entanglement surface is the smaller one. We still keep $l_\phi/z_h$ large, but we also explicitly work with the \textit{(UV-)renormalized} quantities, in which we subtract out the UV divergence,
\begin{align}
A_{HM}^{\text{ren}}(t_0) &= -\frac{2t_0}{z_h} + 2\log z_h + \frac{l_\phi}{z_h} + O\left(\frac{z_h}{l_\phi}\right),\label{regHM}\\
A_{E}^{\text{ren}} &= \frac{2\phi_r}{z_h} + 2\log z_h + \frac{2l_\phi}{z_h} + O\left(\frac{z_h}{l_\phi}\right),
\end{align}
and take $\rho_\infty \to \infty$. The difference in these quantities still matches the difference of the bare areas \eqref{areaDiff}.

First, note that the entropies of both swing surfaces include contributions from the dilaton on the brane, evaluated at the intersection points of the swings with $\phi = 0$. In particular, both the exterior swings' ropes and the Hartman-Maldacena swing's ropes will hit the brane at its boundary, as shown in \eqref{radialRopes}. From the discussion in Appendix \ref{JTgravity}, $\varphi$ has Dirichlet boundary conditions within the brane; denoting its asymptotic values on the left and right boundaries as $\varphi_b^{L,R}/\delta$ with $\delta \to 0$, both entropies will pick up a term proportional to $(\varphi_b^L + \varphi_b^R)/\delta$. This term is seen to be a pure IR divergence in the bulk and cancels out in the difference between the two entropies, so we omit it when writing the renormalized quantities.

Only the two intersection points of the two exterior swings' benches with $\phi = 0$, which we denote as $y_b^L$ and $y_b^R$, will contribute non-divergent terms that appear in the entropy difference,
\begin{equation}
4 G^{(3)} \Delta S(t_0) = \frac{G^{(3)}}{G^{(2)}} \left[\varphi(y_b^L) + \varphi(y_b^R)\right] + \Delta A(t_0) = \kappa + \Delta A(t_0).\label{entDiff}
\end{equation}

We have defined $\kappa$ to encapsulate the dilaton contribution. While $\kappa$ is technically a function of $t_0$ since the locations of the benches change with $t_0$, we will soon find that this dependence is negligible in the large $l_\phi/z_h$ regime. As such, we ignore that dependence for now.

The area difference defined in \eqref{areaDiff} grows in time---the later the time of the radiation region $t_0$, the larger the area difference between the exterior swings and the Hartman-Maldacena swing. However, the Hartman-Maldacena swing is the only one which is non-static. In particular, based on \eqref{HMfinal}, we can see that the Hartman-Maldacena area \textit{decreases} linearly with $t_0$.

The time of the phase transition in the generalized entropy, which we call $t_p$, depends on the interval length $l_\phi$. It occurs when the entropy difference \eqref{entDiff} is $0$, so,
\begin{equation}
t_{p} = -\kappa -\phi_r - \frac{l_\phi}{2} + O\left(\frac{z_h}{l_\phi}\right).\label{phaseT}
\end{equation}

The larger we take $l_\phi$ to be, the earlier the phase transition occurs. In other words, unlike the UV regulator $\rho_\infty$, the IR regulator has an explicit effect on the physics of our configuration.

Next, we define $\tau$ as the time when the renormalized Hartman-Maldacena entropy (which is just the area and does not have any dilaton contribution) reaches $0$. From \eqref{regHM}, we write,
\begin{equation}
\tau = z_h \log z_h + \frac{l_\phi}{2} + O\left(\frac{z_h}{l_\phi}\right).
\end{equation}

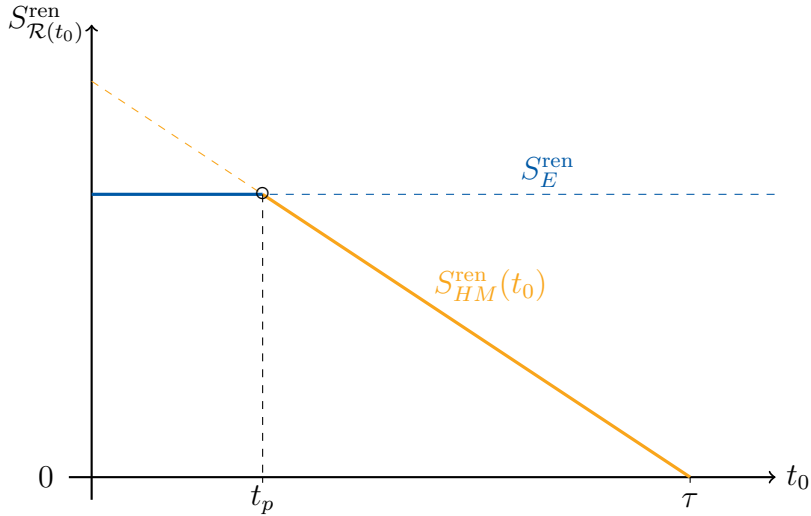
\begin{figure}
\centering
\begin{tikzpicture}[scale=1.5]
\draw[->,thick] (0,-0.2) to (0,4);
\draw[->,thick] (-0.2,0) to (6,0);

\node at (-0.4,4) {$S_{\mathcal{R}(t_0)}^{\text{ren}}$};
\node at (6.2,0) {$t_0$};

\draw[-,blue!30!teal,very thick] (0,2.5) to (1.5,2.5);
\draw[-,dashed,blue!30!teal] (1,2.5) to (6,2.5);

\draw[-,dashed,yellow!20!orange] (0,3.5) to (1.5,2.5);
\draw[-,yellow!20!orange,very thick] (1.5,2.5) to (5.25,0);
\node at (1.5,2.5) {$\circ$};
\draw[-,dashed] (1.5,2.5) to (1.5,0);

\draw[-] (1.5,0) to (1.5,-0.05);
\node at (1.5,-0.2) {$t_p$};

\draw[-] (5.25,0) to (5.25,-0.05);
\node at (5.25,-0.2) {$\tau$};

\node[yellow!20!orange] at (3.5,1.7) {$S_{HM}^{\text{ren}}(t_0)$};
\node[blue!30!teal] at (4,2.7) {$S_{E}^{\text{ren}}$};
\node at (-0.4,0) {$0$};
\end{tikzpicture}
\caption{A sketch of the renormalized entanglement entropy of $\mathcal{R}(t_0)$, $S_{\mathcal{R}(t_0)}^{\text{ren}}$, as a function of time $t_0$, with $S_{HM}^{\text{ren}}(t_0)$ and $S_E^{\text{ren}}$ being the entropies of $A_{HM}^{\text{ren}}(t_0)$ and $A_E^{\text{ren}}$, respectively. The entropy starts at some constant value, then falls after the phase transition in which the entanglement surface becomes the Hartman-Maldacena swing, reaching $0$ at $\tau$.}
\label{figs:entropyCurve}
\end{figure}

The positivity of the excluded divergences allows for the bare area and bare entropy to be positive for all $t_0$. The renormalized entropy curve is depicted in Figure \ref{figs:entropyCurve}.

Observe that this is not the Page curve presented in \cite{Almheiri:2019yqk} and shown in Figure \ref{figs:eternalEntropy}. Instead, it appears to be just the $t < 0$ part of the Page curve, with the Hartman-Maldacena swing decreasing for all time, instead of decreasing, reaching the bifurcation surface at $t = 0$, then increasing. Even more interestingly, if we take $l_\phi/z_h \to \infty$, then both $t_p \to -\infty$ and $\tau \to \infty$. Thus, in the IR limit, we are left with a strictly decreasing entropy curve.

One can ask why the picture is so different from that of AdS/CFT. In fact, Figure \ref{figs:entropyCurve} appears to indicate that the Hawking radiation is in a mixed state, but becomes more pure over time. It is tempting to interpret this as an information paradox. Although the entropy curve gives the appearance of the black hole losing all of its information, the setup is still eternal, and, because we are in the semiclassical limit, there should be no backreaction.

Leaving this puzzle for Section \ref{discussionWarped}, we make another observation with regards to the islands. Note that both the Hartman-Maldacena swing and the exterior swings feature islands. This is because both are truncated by the $\phi = 0$ brane, so both require islands to ``complete" the homology condition of the respective swing surfaces. We compute the endpoints of these islands, confirming that the Hartman-Maldacena swing captures interior degrees of freedom while the exterior swings do not. Our observations here will also prove the earlier statement we made about the time-dependence of $\kappa$ being negligible when $l_\phi/z_h$ is large.

For both candidates, we need to know where the rope emanated from $\mathcal{\phi}_r$ on $\mathcal{R}(t_0)$ intersects with the brane. The rope's affine parameter at this point is \eqref{affineAt0}, and the radial coordinate is \eqref{radialRopes}. Rephrasing both values by keeping only the terms which dominate in the UV limit $\rho_\infty \to \infty$, we write,
\begin{align}
\lambda_0 &\to z_h \rho_\infty e^{-2\phi_r/z_h},\\
z_-(\lambda_0) &\to 0\ \ (\text{from above}).
\end{align}

Thus the intersection point lies just within the interface between the brane and the bath, \textit{i.e.} approaching\footnote{We are careful to note that, even in the UV limit, the intersection point does not lie \textit{on} the AdS$_2$ boundary, as this would violate the boundary conditions of the setup. The intersection point must be strictly in the AdS$_2$ bulk.} the conformal boundary of the induced 2-dimensional black hole. Furthermore, we use \eqref{ropesAS} to write the time coordinate as,
\begin{equation}
t_-(\lambda_0) \to -\frac{z_h}{2}\log\left(e^{-2\phi_r/z_h}\right) + t_0 = t_0 + \phi_r.
\end{equation}

So, on both sides of the brane's black hole, the islands corresponding to both candidate swing surfaces start at $t = t_0 + \phi_r$ and $z \to 0$.

\begin{figure}
\centering
\begin{tikzpicture}[scale=2.5]
\draw[-,draw=none,fill=black!10] (1,1) to (1,-1) to (0,0) to (1,1); 
\draw[-,draw=none,fill=black!10] (-1,1) to (-1,-1) to (0,0) to (-1,1);

\draw[-,draw=none,fill=violet!10] (-1,1) to[bend right] (1,1) to (0,0) to (-1,1); 
\draw[-,draw=none,fill=violet!10] (1,-1) to[bend right] (-1,-1) to (0,0) to (1,-1);

\draw[-,decoration = {zigzag,segment length = 1mm, amplitude = 0.25mm},decorate] (-1,1) to[bend right] (1,1);
\draw[-] (1,1) to (1,-1);
\draw[-,decoration = {zigzag,segment length = 1mm, amplitude = 0.25mm},decorate] (1,-1) to[bend right] (-1,-1);
\draw[-] (-1,-1) to (-1,1);

\draw[-,dashed,color=red] (-1,1) to (1,-1);
\draw[-,dashed,color=red] (1,1) to (-1,-1);

\draw[-] (1,-1) to (2.25,0) to (1,1) to (1,-1);
\draw[-] (-1,-1) to (-2.25,0) to (-1,1) to (-1,-1);

\node at (1.35,0.125) {\textcolor{black!25!lime}{\footnotesize$\bullet$}};
\node at (2.025,0.125) {\textcolor{black!25!lime}{\footnotesize$\bullet$}};

\node at (-1.35,0.125) {\textcolor{black!25!lime}{\footnotesize$\bullet$}};
\node at (-2.025,0.125) {\textcolor{black!25!lime}{\footnotesize$\bullet$}};

\draw[-,black!25!lime,,very thick] (1.35,0.125) to (2.025,0.125);
\draw[-,black!25!lime,,very thick] (-1.35,0.125) to (-2.025,0.125);

\node at (3.375/2,0) {\textcolor{black!25!lime}{$\mathcal{R}^R(t_0)$}};
\node at (-3.375/2,0) {\textcolor{black!25!lime}{$\mathcal{R}^L(t_0)$}};

\draw[-] (0.95,0.125) to (1.05,0.125);
\node at (1.15,0.125) {$t_0$};
\draw[-] (-0.95,0.125) to (-1.05,0.125);
\node at (-1.15,0.125) {$t_0$};

\draw[-] (0.95,0.475) to (1.05,0.475);
\node at (1.33,0.475) {$t_0 + \phi_r$};
\draw[-] (-0.95,0.475) to (-1.05,0.475);
\node at (-1.33,0.475) {$t_0 + \phi_r$};

\draw[-] (0.95,0.8775) to (1.05,0.8775);
\draw[-,draw=none,fill=white] (1.075,0.95) to (1.075,0.8) to (1.225,0.8) to (1.225,0.95) to (1.075,0.95);
\node at (1.15,0.8775) {$t_*$};

\draw[-] (-0.95,0.8775) to (-1.05,0.8775);
\draw[-,draw=none,fill=white] (-1.075,0.95) to (-1.075,0.8) to (-1.225,0.8) to (-1.225,0.95) to (-1.075,0.95);
\node at (-1.15,0.8775) {$t_*$};

\draw[-,blue!30!teal,very thick] (1-0.005,0.475+0.025) to (0.955+0.0025,0.8775);
\node at (0.955,0.8775) {\footnotesize\textcolor{blue!30!teal}{$\bullet$}};
\draw[-,blue!30!teal,very thick] (-1+0.005,0.475+0.025) to (-0.955-0.0025,0.8775);
\node at (-0.955,0.8775) {\footnotesize\textcolor{blue!30!teal}{$\bullet$}};

\draw[-,yellow!20!orange,very thick] (-1+0.0275,0.475) to (-0.475,0.475) .. controls (0,0.375) .. (0.475,0.475) to (1-0.0275,0.475);

\draw[-,draw=blue!30!teal,thick] (1+0.0275,0.479) arc (0:180:0.0275); 
\draw[-,draw=yellow!20!orange,thick] (1+0.0275,0.479) arc (0:-180:0.0275); 

\draw[-,draw=blue!30!teal,thick] (-1-0.0275,0.479) arc (180:0:0.0275); 
\draw[-,draw=yellow!20!orange,thick] (-1-0.0275,0.479) arc (-180:0:0.0275);

\node at (0.85,0.65) {\textcolor{blue!30!teal}{$\mathcal{I}_E^R$}};
\node at (-0.85,0.65) {\textcolor{blue!30!teal}{$\mathcal{I}_E^L$}};

\node at (0.6,0.3) {\textcolor{yellow!20!orange}{$\mathcal{I}_{HM}$}};
\end{tikzpicture}
\caption{The $2$-dimensional system, consisting of the black hole on the $\phi = 0$ brane coupled to the bath on the conformal boundary. The radiation region is depicted in green and is the union of two disconnected pieces. The blue intervals comprise a sketch of the exterior islands, whereas the orange interval is a sketch of the Hartman-Maldacena island. We have taken $\rho_\infty \to \infty$, so both intervals \textit{approach} the conformal boundary (a phenomenon represented by open dots) and thus include \textit{near-boundary} degrees of freedom.}
\label{figs:penroseIslands}
\end{figure}
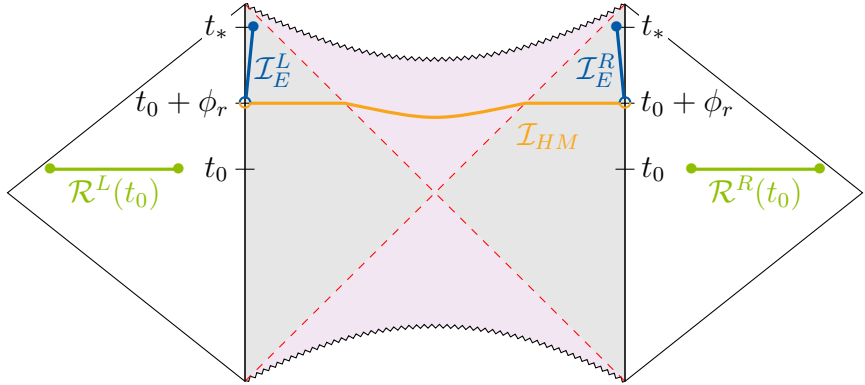

The only intersection points of the full Hartman-Maldacena swing with the $\phi = 0$ brane are precisely these points. Thus, the \textit{Hartman-Maldacena island}, $\mathcal{I}_{HM}$, is the interval connecting the two. This is shown in Figure \ref{figs:penroseIslands}.

Each exterior swing, however, intersects $\phi = 0$ both at the point described above and at a second point just outside of the horizon. Specifically, using \eqref{benchAS}, we immediately have that the radial coordinate of this other intersection is,
\begin{equation}
z_* = z_h\sqrt{\frac{2}{1 + \coth\left(\frac{l_\phi}{2z_h}\right)}} = z_h + O\left(\frac{z_h}{l_\phi}\right) \xrightarrow{l_\phi/z_h \to \infty} z_h.
\end{equation}

In other words, for either of the exterior swings, the deeper intersection point approaches the horizon in the IR limit, but never enters the interior. Note that we may find the time coordinate $t_*$ of these endpoints, as well. From \eqref{benchLC} and the fact that $\phi = 0 \implies v_* = -u_*$, where $v_*$ and $u_*$ are the lightcone coordinates at this point, we deduce,
\begin{equation}
t_* = \frac{u_* - v_*}{2} = u_* = \frac{u_+ + u_-}{2}.
\end{equation}

We still have \eqref{uPMRad}, which yields,
\begin{equation}
t_* = t_0 + \phi_r + \frac{l_\phi}{2} \xrightarrow{l_\phi/z_h \to \infty} \infty.
\end{equation}

These \textit{exterior islands} $\mathcal{I}_E = \mathcal{I}_E^R \cup \mathcal{I}_E^L$ are also shown in Figure \ref{figs:penroseIslands}. That $z_* \to z_h$ and $t_* \to \infty$ in the IR limit validates our earlier statement that the dependence of $\kappa$ on $t_0$ may be neglected for large $l_\phi/z_h$, since we deduce that, regardless of $t_0$, the deeper endpoint of the exterior islands is \textit{fixed} to be near the horizon in the IR limit.

Additionally, the endpoints of the exterior islands appear to be timelike, as opposed to spacelike like the Hartman-Maldacena island. This only happens because the exterior swings, which do not lie on a single time slice, hit the brane at different times. That the separation is timelike is easily seen in the IR limit, where the differences in the $t$ and $z$ coordinates between the near-boundary endpoint and near-horizon endpoint are respectively,
\begin{equation}
\Delta t = t_* - (t_0 + \phi_r) \xrightarrow{l_\phi/z_h \to \infty} \infty,\ \ \Delta z = z_* \xrightarrow{l_\phi/z_h} z_h.
\end{equation}

So taking stock, we have done the following: defining constant-time radiation regions in the bath, we use double holography to compute their entanglement entropies via the swing surface prescription. In doing so, we find early-time islands which are timelike. That these exterior islands are timelike is a statement about how early-time information is stored in our setup---for $t_0 < t_p$, a single observer has access to the radiation collected in one side of the bath.

To elaborate on the physics of this point, we first note that each connected exterior island $\mathcal{I}_E^{R,L}$ (approaching the boundary at time $t_0 + \phi_r$) forms in the 3-dimensional bulk a single connected entanglement wedge $\Sigma_E^{R,L}$ with the radiation collected in the respective bath $\mathcal{R}^{R,L}(t_0)$ and the appropriate exterior swing. Thus, each exterior island $\mathcal{I}_E^{R,L}$ is redundant with the corresponding radiation region $\mathcal{R}^{R,L}(t_0)$---a stronger statement than $\mathcal{I}_E$ being redundant with the full radiation region.

\textbf{A note of caution:} The timelike nature of the islands makes it difficult to assign them a standard Hilbert-space description. This challenges the premise that the radiation region is encoded within the island region, or vice versa. It is possible that there is some effective and finer notion of this encoding at each time slice on the timelike island region, with the spacelike radiation region, which entanglement entropy itself cannot decipher. However, we have no sharp and precise statements to make on how this could be true.

With the open and unsettled issues mentioned above, technically we have simply combined various methods, each of which have independent support. The puzzling aspects of our results could stem from a hitherto unknown limitation or inconsistency in those methods, or it could be solved by a better understanding of the island description in more general situations. At this point, we are not sure how to better think about them and we leave these issues for future.\footnote{We thank the Referee for extremely useful feedback on the manuscript and bringing these questions to the spotlight.}

In the following section, leaving the above issues aside, we compare the above statements to the analogous ones in AdS/CFT, in which the RT prescription is what calculates the entanglement entropy. Furthermore, we discuss the how our model provides insight into the relationship between islands and unitarity, as well as the role of the IR divergence.

\subsection{Differences from AdS/CFT}\label{discussionWarped}

In Appendix \ref{islandsAdSCFT}, we review the analogous calculations for the static planar BTZ in AdS$_3$/CFT$_2$, using the arguments of \cite{Geng:2020qvw} to arrive at the statements of \cite{Almheiri:2019yqk}. Following this procedure, which roughly relies on the same logic as our analysis using swing surfaces, we ultimately arrive at three conclusions:
\begin{itemize}
\item[(1)] the entanglement entropy of the radiation region follows the eternal Page curve\footnote{We emphasize that, by time-reversal symmetry, the plot for negative times is the mirror image with respect to vertical axis. In other words, the entanglement entropy is also saturated for sufficiently negative times.} of Figure \ref{figs:eternalEntropy},
\item[(2)] there are islands (specifically going behind the horizon) only for the exterior surface, but no islands for the Hartman-Maldacena surface,
\item[(3)] there is no field-theoretic IR divergence, since, with the RT prescription, we may take the boundary interval to start at $\phi_r$ and be infinite in length.
\end{itemize}

All three of these conclusions change when the matter sector is a holographic WCFT$_2$. Mathematically, these differences emerge by virtue of using the swing surface prescription. We elaborate on each of these difference from a physical perspective.

Starting with (1), we have already found the entropy curve, depicted in Figure \ref{figs:entropyCurve}. Essentially, both the phase transition and saturation \textit{only} occur at some early time (past infinity in the IR limit). After that, the curve simply decreases, indicating that the Hawking radiation is becoming more and more pure.

We have already described this as having the appearance of an information paradox, but recall that holographic WCFT$_2$ is actually \textit{non-unitary}. From a field theory perspective, we posit that it is this non-unitarity which leads to an ever-decreasing Page curve, even though the setup is supposed to be eternal and the Hawking radiation in equilibrium with the black hole.

Additionally, observe that Figure \ref{figs:entropyCurve} is not even symmetric under time-reversal, even though the bulk geometry is. This indicates that boundary state living in the WCFT$_2$ is not invariant under time-reversal.

For (2), as shown in Figure \ref{figs:penroseIslands}, both the Hartman-Maldacena swing and the exterior swings feature islands anchored to the boundary at different time slices from $\mathcal{R}(t_0)$. Physically, this means that, no matter what time at which we compute the entanglement of the Hawking radiation with the black hole, we will always find that there are redundant degrees of freedom in the gravitating region. Additionally, the exterior islands are \textit{strictly} in the exterior regions of the black hole, while the Hartman-Maldacena islands cross the interior. That the ``late-time" islands cross the interior is an interesting common point between the warped story and the AdS/CFT story. However, regarding the islands' locations, that is where the similarities appear to end.

That being said, however, it is interesting that the islands emerge at all. Essentially, the warped model tells us that the idea of the island rule and redundant degrees of freedom existing within the 2-dimensional gravitating region does not necessarily imply unitarity. Rather, redundancy is a more general idea. In conjunction with past work, we deduce that the relationship between unitarity and islands is rather subtle---\cite{Alishahiha:2020qza} considers two higher-derivative, non-unitary gravitational theories and finds unitary Page curves and late-time islands, while we take the \textit{matter} to be non-unitary in the course of deviating from this story.

It is perhaps worth emphasizing that the presence of the islands even for the Hartman-Maldacena surface is strongly suggestive of a deeper connection between the geometric notion of islands and the intrinsic non-local nature of the quantum interaction. In AdS/CFT, the resolution of the unitary Page curve may be related to the non-locality of quantum gravity interactions, which are captured in terms of the islands but are unknown to the Hartman-Maldacena surface. In our case, however, the WCFT$_2$ is inherently non-local and it is tempting to conclude that, therefore, the Hartman-Maldacena surface cannot evade islands. To further establish a precise connection between non-locality and the presence of islands, one at least needs to understand such models in a more controlled manner. For example, one may attempt to construct a set-up in which a tunable parameter smoothly interpolates between locality and non-locality, with the hope of attaining an ``island-free" entanglement surface in the local limit. We leave this to future work.

Lastly, regarding (3), we again reiterate that the physics depends on the field-theoretic IR cutoff $l_\phi$ which corresponds to the size of the radiation region. Specifically,
\begin{equation}
\frac{l_\phi}{z_h} \to \infty \implies \begin{cases}
t_p \to -\infty,\\
\tau \to \infty,\\
S^{\text{ren}}_{\mathcal{R}(-\infty)} \to \infty.
\end{cases}
\end{equation}

Thus, in the IR limit, the non-unitarity of the entropy curve becomes even more evident. Not only is there no phase transition, but the entropy is \textit{always} strictly decreasing, with the renormalized entanglement entropy always being positive (albeit infinite, since we have not renormalized with respect to the IR divergence).

In particular, note that these IR divergences are \textit{linear}. \cite{Castro:2015csg} makes the point that the extra linear term seen in the warped entanglement entropy expression\footnote{The linear term is an ``extra" one when comparing the WCFT$_2$ entanglement entropy to the CFT$_2$ entanglement entropy.} are a manifestation of its non-relativistic nature, stemming from the theory's lack of Lorentz invariance. As it is these very terms which give rise to the divergences in the physical quantities above, one may wonder about how loss of symmetry in the field theory may be related to these physical divergences. We leave this to future consideration.

However, our quantitative results are still expanded around the IR limit, as well. One could ask what happens when $l_\phi/z_h$ is \textit{not} sufficiently large for such expressions to be good. Then, the time-dependence of $\kappa$ is no longer suppressed, and the result \eqref{phaseT} would be modified so as to be dependent on the profile of the JT-like dilaton.

While AdS$_3$/WCFT$_2$ and the swing surface prescription ultimately give a radically different result from AdS$_3$/CFT$_2$ despite the logic being similar (see Appendix \ref{islandsAdSCFT} for details), the moral of the story is that the emergence of entanglement islands are ultimately a ``well-behaved" phenomenon, in that they need not always impose unitarity. In other words, islands are not \textit{too} strong. We may claim that when an entropy curve is unitary and thus described by a Page curve, this is a physical phenomenon informed by the details of the theory. In conjunction with the results of \cite{Alishahiha:2020qza}, we deduce that islands are \textit{only} a necessary ingredient to obtain a Page curve, but are not powerful enough to overcome all non-unitary variations of the AdS/CFT doubly holographic models.

\section{Conclusions}\label{conclusion}

To summarize, we have used the recently-proposed swing surface prescription\cite{Apolo:2020bld,Apolo:2020qjm} for non-AdS holography in order to compute the entanglement entropy of Hawking radiation in a doubly holographic, eternal model using AdS$_3$/WCFT$_2$. Our primary goal is two-fold: to clarify whether the island prescription is ``too powerful” and always reproduces a unitary Page curve even when the system is non-unitary, and to explore further the nature of the non-locality of the island rule.

By choosing the WCFT$_2$ matter, we explicitly couple gravity to a non-unitary, non-local system. We find that the corresponding Page curve reflects this non-unitarity, and that islands are always present. We therefore interpret this as evidence in support of the following:
\begin{itemize}
\item[(1)] the island prescription actually knows about whether the microscopic description is indeed unitary or non-unitary,

\item[(2)] the island prescription also responds to the microscopic non-locality of the system.
\end{itemize}

The two points are primarily based on plausibility. In other words, we are suggesting that one should perhaps take the above coincidences seriously, and the islands may carry further fine-grained details of the gravitational description including the matter field to which it is coupled. In brief, the pathologies of the class of models seem to be well-captured by the island rule in the gravitational dual.

In the course of our work, we have learned that entanglement islands are a rather necessary phenomenon in gravitating systems, emerging even when the accompanying entropy curve is not unitary. There are a number of directions in which one may proceed from here.

\begin{itemize}
\item  We stress that we have only discussed a \textit{single} alternative doubly holographic model to that used in \cite{Almheiri:2019hni}. The work on swing surfaces, however, encompasses other types of non-AdS holography using 2-dimensional field theories. By restricting to holographic models in which the swing surface proposal holds, one could perform a similar analysis in a much broader class of situations, including WAdS$_3$/WCFT$_2$ (which changes the black hole to being non-Einsteinian) and flat$_3$/BMFT$_2$\cite{Bagchi:2014iea,Jiang:2017ecm}.

\item One may consider non-AdS doubly holographic models in higher dimensions. For instance, one may consider non-relativistic Lipshitz theories, for which entanglement wedges have been studied\cite{Gentle:2015cfp,Cheyne:2017bis}. Note that \cite{Gentle:2015cfp} mentions the necessity of ropes (or, in this case, null sheets) in the full entanglement surface.

\item Within the confines of AdS$_3$/WCFT$_2$, one may also consider whether there are islands for radiation emitted by a static AdS$_3$ vacuum. Since this is another vacuum solution, the swing surfaces computed by \cite{Apolo:2020bld} should work. Note, however, that factors of $i$ may need to be carefully considered.

\item Also within AdS$_3$/WCFT$_2$, one could consider how to source black hole evaporation in the semiclassical limit. However, this would not longer be a vacuum solution, so the swing surfaces would need to be found.

\item Although a given WCFT$_2$ is expected to be non-local, we never explicitly used this property in our work. It would be interesting to study islands in non-local theories. In order to properly probe any potential effects, our work further indicates that one would need to employ unitary theories. Another route is to construct a system which smoothly interpolates between locality and non-locality so as to see if some entanglement surfaces featuring islands in the latter become ``island-free" in the former.

\item Our calculations were performed completely holographically in the 3-dimensional braneworld system, but in principle it should be possible to perform these calculations in the other systems. For example, in the 2-dimensional gravity + bath system, one could compute the entanglement entropy by first generalizing the expression in \cite{Detournay:2012pc} to curved backgrounds. This should provide a consistency check with our holographic calculations. However this would require knowing more about the JT-like action.

\item A field-theoretic IR divergence is itself dual to a UV divergence in the gravity theory---in other words, a source. While we found a field-theoretic IR divergence in our work, it is unclear what type of source it may correspond to in the bulk. Furthermore, from the field-theoretic perspective, the origin of this IR divergence is also unclear. One possible way of conceptualizing this source is as something which breaks the full Lorentz invariance of AdS$_3$, tying it to the non-relativistic nature of the boundary WCFT$_2$. More generally, one may ask about the possible role of IR divergences in non-relativistic field theories, particularly holographic ones.
\end{itemize}

\section*{Acknowledgments}

We thank Luis Apolo for thorough feedback and comments on the manuscript. We also thank Hao Geng, Hongliang Jiang, Andreas Karch, Wei Song, and Yuan Zhong for useful discussions. EC and SS are supported by National Science Foundation (NSF) Grant No. PHY-1820712. SS is also supported by NSF Grant No. PHY–1914679. AK and AKP are supported by the Department of Atomic Energy, and the Council of Scientific \& Industrial Research (CSIR), Gov't of India. AKP is supported by CSIR Fellowship No. 09/489(0108)/2017-EMR-I.

\begin{appendices}
\section{JT Gravity Coupled to CFT$_2$ Matter}\label{JTgravity}

We briefly review JT gravity coupled to conformal matter, starting at the classical level, then promoting everything to the semiclassical level. This is discussed in \cite{Almheiri:2019psf} in the context of quantum extremal surfaces. Additionally, it is one of the descriptions for the doubly holographic model of \cite{Almheiri:2019hni}. When coupling JT gravity to a CFT, classically the matter's stress tensor ends up coupled to the dilaton, while the background has a fixed Ricci curvature. Since we neglect any quantum backreaction on the metric in the semiclassical approximation, the background remains fixed, and occurrences of the stress tensor are replaced with its expectation value.

We use the action in \cite{Harlow:2018tqv}.\footnote{The action presented by \cite{Harlow:2018tqv} includes a holographic renormalization meant to keep the action finite on relevant classical configurations.} In particular, we show that the action is extremized for configurations involving fixed AdS$_2$ backgrounds, with the matter's stress tensor being related to the dilaton by an additional set of on-shell constraints. We also specify boundary conditions on the metric and the dilaton. When considering the JT-like gravity coupled to WCFT$_2$ in the main text, we assume the same sort of classical configuration, with the same boundary conditions and coupling to matter for the dilaton.

The action of JT gravity itself consists of two separate parts,
\begin{equation}
I_{JT}[g_{ij}^{(2)},\varphi] = I_{T}[g_{ij}^{(2)}] + I_{G}[g_{ij}^{(2)},\varphi],
\end{equation}
where these terms are defined as,
\begin{align}
I_{T}[g_{ij}^{(2)}] &= \frac{\varphi_0}{16\pi G^{(2)}} \left(\int_{\mathcal{M}} d^2 x\sqrt{-g}R + 2\int_{\partial\mathcal{M}} dx \sqrt{|\gamma|}K\right),\\
I_{G}[g_{ij}^{(2)},\varphi] &= \frac{1}{16\pi G^{(2)}}\left[\int_{\mathcal{M}} d^2 x \sqrt{-g}\varphi\left(R + \frac{2}{\ell^2}\right) + 2\int_{\partial\mathcal{M}} dx \sqrt{|\gamma|}\varphi(K-1)\right].\label{dynamicalAct}
\end{align}

Here, $g_{ij}^{(2)}$ is the background metric and $\varphi$ is the dynamical dilaton. We also have couplings $G^{(2)}$ and $\varphi_0 \gg \varphi$. However, although $I_G$ is a dynamical term, $I_{T}$ is actually topological. This is seen by the Gauss-Bonnet theorem in the Euclidean sector; for a 2-dimensional orientable, Riemannian manifold $\mathcal{M}_E$ with Euler characteristic $\chi(\mathcal{M}_E)$,
\begin{equation}
\int_{\mathcal{M}_E} d^2 x\sqrt{g_E}R + 2\int_{\partial\mathcal{M}_E} dx \sqrt{\gamma_E}K = 4\pi\chi(\mathcal{M}_E).
\end{equation}

Thus, the Euclideanized $I_T$ in the JT gravity action is,
\begin{equation}
I^E_{T} = -\frac{\varphi_0 \chi(\mathcal{M}_R)}{4G^{(2)}} = \frac{\varphi_0}{4G^{(2)}}(2g + b - 2),
\end{equation}
where $g$ is genus and $b$ is the number of boundaries.

In the path integral, any term of the form $\exp(-I_T^E)$ corresponding to a configuration with large $g$ or large $b$ will be exponentially suppressed. Consequently, in the approximation for which we consider the leading-order term, we take $g = 0$ and $b = 1$. As mentioned in \cite{Almheiri:2019psf}, this means that, semiclassically, the topological term yields the following leading-order contribution to the entropy,
\begin{equation}
S_T \approx \log \exp(\frac{\varphi_0}{4G^{(2)}}) = \frac{\varphi_0}{4G^{(2)}}.
\end{equation}

Now, in finding the classical configurations for the 2-dimensional bulk in \eqref{2dSystem}, we can neglect the variation of $I_T$ (since it is topological). Furthermore, we apply the Dirichlet boundary conditions in \cite{Harlow:2018tqv,Almheiri:2019psf} to fix the boundary metric and the boundary value of the dilaton,
\begin{equation}
\gamma_{uu}|_{\partial\mathcal{M}} = \frac{1}{\delta^2},\ \ \varphi|_{\partial\mathcal{M}} = \frac{\varphi_b}{\delta},\label{dilatonBC}
\end{equation}
taking $\delta \to 0$ and $\varphi_b > 0$ finite. With these boundary conditions, we can neglect the variation of the boundary terms in the action.

We now focus on varying just the bulk parts of $I_G$ and $I_{W}$. First, varying by $\varphi$, we find that,
\begin{equation}
\frac{\delta}{\delta \varphi}\left(I_{G} + I_{W}\right) = \frac{1}{16\pi G^{(2)}} \sqrt{-g}\left(R + \frac{2}{\ell^2}\right).\label{eomBackground}
\end{equation}

Classically, we thus have that the scalar curvature is fixed. Furthermore, this extends to the semiclassical regime because we do away with any quantum backreaction in the metric. Specifically, the background must be locally AdS$_2$,
\begin{equation}
R = -\frac{2}{\ell^2}.\label{eomBackground2}
\end{equation}

Next, we vary with respect to the bulk metric. We use the standard definition for the matter stress tensor,
\begin{equation}
T_{ij} = -\frac{2}{\sqrt{-g}} \frac{\delta I_{W}}{\delta g^{ij}}.
\end{equation}

Upon computing the variation of \eqref{dynamicalAct}, we find that,
\begin{align}
\frac{\delta}{\delta g^{ij}}\left(I_{G} + I_{W}\right)
=\ &\frac{\sqrt{-g}}{16\pi G^{(2)}}\left[-\frac{1}{2}g_{ij}\left(R + \frac{2}{\ell^2}\right)\varphi + R_{ij}\varphi - \nabla_i \nabla_j \varphi + g_{ij}\nabla^2 \varphi\right]\nonumber\\
&- \frac{\sqrt{-g}}{2} T_{ij}.
\end{align}

We can apply \eqref{eomBackground2} to eliminate the first term in the brackets. Additionally, setting this to $0$ yields the following classical stress tensor,
\begin{equation}
8\pi G^{(2)} T_{ij} = (R_{ij} + g_{ij}\nabla^2 - \nabla_i \nabla_j)\varphi.\label{stress1}
\end{equation}

We conclude by showing that this results in the stress tensor equations of \cite{Almheiri:2019psf}. As in their work, consider a local AdS$_2$ patch of the background in lightcone coordinates,
\begin{equation}
\frac{ds^2}{\ell^2} = -\frac{4dx^+ dx^-}{(x^+ - x^-)^2}.
\end{equation}

In these coordinates, the three independent components of \eqref{stress1} are,
\begin{align}
8\pi G^{(2)}T_{+-} &= \partial_+ \partial_- \varphi + \frac{2}{(x^+ - x^-)^2}\varphi,\\
8\pi G^{(2)}T_{++} &= -\frac{1}{(x^+ - x^-)^2}\partial_+ \left[(x^+ - x^-)^2 \partial_+ \varphi\right],\\
8\pi G^{(2)}T_{--} &= -\frac{1}{(x^+ - x^-)^2}\partial_- \left[(x^+ - x^-)^2 \partial_- \varphi\right].
\end{align}

Thus, \eqref{stress1} is consistent with \cite{Almheiri:2019psf}. Furthermore, in the semiclassical approximation, we write,
\begin{equation}
8\pi G^{(2)} \expval{T_{ij}} = (R_{ij} + g_{ij}\nabla^2 - \nabla_i \nabla_j)\varphi.\label{stress2SC}
\end{equation}

\section{Hartman-Maldacena Area in the Interior}\label{areaHM}

The analysis performed by \cite{Hartman:2013qma} uses a $(d+1)$-dimensional metric of the form,
\begin{equation}
ds^2 = -g(\rho)^2 d\tilde{t}^2 + f(\rho)^2 d\vec{x}^2 + d\rho^2,\label{metHM}
\end{equation}
where $\vec{x} = (x^1,...,x^{d-1})$ and the functions $g$ and $h$ are,
\begin{equation}
g(\rho) = f(\rho)\tanh\left(\frac{d}{2}\rho\right),\ \ f(\rho) = \frac{2}{d}\left[\cosh\left(\frac{d}{2}\rho\right)\right]^{2/d}
\end{equation}

The Hartman-Maldacena surface sits at a constant $x^1$, and its $\rho$-coordinate is parameterized as $\rho(\tilde{t})$. Furthermore, in the exterior, it is anchored to a boundary region at $\tilde{t} = \tilde{t}_b$. Denoting the derivative of $\rho(\tilde{t})$ by $\dot{\rho}$, the induced metric on the surface is,
\begin{equation}
ds^2|_{HM} = [-g(\rho)^2 + \dot{\rho}^2]d\tilde{t}^2 +f(\rho)^2 \sum_{i=2}^{d-1} (dx^{i})^2.
\end{equation}

Thus, we have that the area is,
\begin{align}
A
&= \int dx^2 \cdots dx^{d-1} \int d\tilde{t}\, f(\rho)^{d-2} \sqrt{-g(\rho)^2 + \dot{\rho}^2}\nonumber\\
&= V_{d-2} \int d\tilde{t}\, f(\rho)^{d-2} \sqrt{-g(\rho)^2 + \dot{\rho}^2},\label{areaFunc}
\end{align}
where $V_{d-2}$ is the volume of the transverse space $(x^2,...,x^{d-1})$.

In the interior, the signatures of $\tilde{t}$ and $\rho$ switch, so $\rho$ becomes a timelike coordinate. In other words, $\rho$ is imaginary in the interior. Bearing this in mind, \cite{Hartman:2013qma} extremizes \eqref{areaFunc}, defining along the way,
\begin{equation}
a(\rho) = -ig(\rho)f(\rho)^{d-2}.\label{aFunc}
\end{equation}

For late boundary time $\tilde{t}_b$ (or by time-reversal symmetry, early boundary time $-\tilde{t}_b$), \cite{Hartman:2013qma} finds the area of the portion of the extremal surface in the interior to be,
\begin{equation}
A_{\text{int}} = 2V_{d-2}a_m \tilde{t}_b,\label{areaInt}
\end{equation}
where $a_m$ is the maximum value of the function \eqref{aFunc} acquired in the interior, when $\rho$ is imaginary. The exterior portion of the surface is on the $\tilde{t} = \tilde{t}_b$ slice of the geometry, so it is constant in time.

For $d = 2$, we set $x^1 = \tilde{\phi}$ and rewrite \eqref{metHM} as,
\begin{equation}
ds^2 = -\sinh^2 \rho\,d\tilde{t}^2 + \cosh^2\rho\,d\tilde{\phi}^2 + d\rho^2.\label{metHM2}
\end{equation}

Then, we apply the coordinate transformation,
\begin{equation}
\tilde{t} = \frac{t}{z_h},\ \ \tilde{\phi} = \frac{\phi}{z_h},\ \ \rho = \Sech^{-1}\left(\frac{z}{z_h}\right),
\end{equation}
under which \eqref{metHM2} becomes the metric of the exterior \eqref{btzAS}. As there are no additional transverse coordinates besides $\tilde{\phi}$,
\begin{equation}
V_0 = 1.
\end{equation}

Furthermore, in the interior, if we define $\rho = i\kappa$, $\kappa \in \mathbb{R}$, then \eqref{aFunc} becomes,
\begin{equation}
a(\kappa) = -i\sinh(i\kappa) = \sin\kappa.
\end{equation}

So, as $a_m$ is the maximum, we deduce that $a_m = 1$. We conclude that, in AdS-Schwarzschild time, the interior contribution to the area of the Hartman-Maldacena surface anchored to a late-time $t = t_b$ slice in the exterior is,
\begin{equation}
A_{\text{int}} = \frac{2t_b}{z_h}.
\end{equation}

\section{Islands in AdS/CFT}\label{islandsAdSCFT}

We briefly review the island story for the eternal BTZ, but in AdS/CFT. \cite{Geng:2020qvw} specifically performs calculations in an AdS$_5$ planar black hole containing a probe brane, but the general argument is similar for the 3-dimensional planar BTZ. This time, we use the RT prescription, and the radiation region consists of two intervals of the form,
\begin{equation}
\{t_0\} \times [\phi_r,\infty).
\end{equation}

\begin{figure}
\centering
\begin{tikzpicture}[scale=1.5]
\draw[->] (0,2) to (2+2,2);
\draw[-,very thick,blue] (0,2) to (0,0);

\draw[->] (0,2.3) to (0.5,2.3);
\node at (0,2.5) {$\phi$};

\draw[->] (-0.3,2) to (-0.3,1.5);
\node at (-0.5,1.5) {$z$};

\node at (0,-0.2) {\textcolor{blue}{$\phi = 0$}};

\draw[red,-,dashed] (0,0.2) to (2+2,0.2);

\node at (2+2.6,0.2) {\textcolor{red}{$z = z_h$}};
\node at (2+2.6,2) {$z = 0$};

\draw[-,yellow!20!orange,very thick] (1.5,2) to (1.5,0.2);
\draw[-,blue!30!teal,very thick] (1.5,2) to[bend left] (0,1.225);
\draw[-,blue!30!teal] (0.1,1.225) to (0.1,1.325) to (0,1.325);

\draw[->,very thick,black!25!lime] (1.5,2) to (4,2);
\node at (2.625,2.2) {\textcolor{black!25!lime}{$\mathcal{R}(t_0)$}};

\draw[-,black!25!lime,thick] (1.6,2.2) to (1.5,2.2) to (1.5,1.8) to (1.6,1.8);
\node at (1.5,2.4) {\textcolor{black!25!lime}{$\phi_r$}};

\node[color=blue] at (0,2) {$\bullet$};
\end{tikzpicture}
\caption{The candidate RT surfaces of an infinite-length radiation region confined to a single exterior region of the two-sided planar BTZ at $t = t_0$. The orange line entering the horizon is the Hartman-Maldacena surface, whereas the blue arc which ends perpendicularly on the brane is the island-producing exterior surface.}
\label{figs:braneworldCFT}
\end{figure}
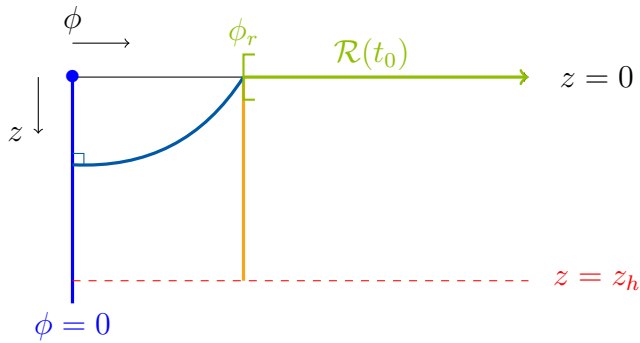

In other words, we need not bother with a field-theoretic IR regulator. The candidate RT surfaces in one of the exterior regions are shown in Figure \ref{figs:braneworldCFT}.

It is straightforward to calculate the area of the Hartman-Maldacena surface in a single exterior region by using \eqref{btzAS}. However, note that it exhibits a logarithmic boundary divergence, as is standard for AdS$_3$ geometries.
\begin{equation}
\tilde{A}_{HM} = \int_0^{z_h} \frac{dz}{z\sqrt{h(z)}}.
\end{equation}

We parameterize the other RT candidate, which we call the \textit{exterior surface}, as $\phi(z)$. By definition, it must extremize the area functional,
\begin{equation}
\tilde{\mathcal{A}}[\phi'] = \int \frac{dz}{z}\sqrt{\frac{1}{h(z)} + \phi'(z)^2}.
\end{equation}

The resulting variational derivative gives us two conditions: (1) any extremal surface hits the brane \textit{orthogonally} (as shown in Figure \ref{figs:braneworldCFT}) and (2) the trajectory must satisfy,
\begin{equation}
\phi'(z) = -\frac{z}{\sqrt{h(z)(z_*^2 - z^2)}},\label{derivPhi}
\end{equation}
where $z_*$ is the \textit{turnaround point} at which $1/\phi'(z_*) = 0$. This turnaround point is an integration constant, defining a class of surfaces which satisfy the bulk equations of motion but need not satisfy the boundary conditions, and taking $z_* \to \infty$ yields the Hartman-Maldacena surface.

While there is \textit{formally} a positive branch in addition to \eqref{derivPhi}, we can show that $\phi' < 0$ for any extremal trajectory going from the conformal boundary $\phi > 0$ to the probe brane $\phi = 0$. Specifically, the orthogonality boundary condition at the brane indicates that $\phi(z_*) = 0$, \textit{i.e.} the turnaround point lies on the probe brane. As such, along the extremal trajectory from the conformal boundary to the probe, $z$ only ever increases while $\phi$ decreases.

To compute the area of the exterior surface, we need to compute the turnaround point in terms of boundary and geometric parameters. By integrating \eqref{derivPhi} over $z \in (0,z_*)$, we have that,
\begin{equation}
0 = \phi(z_*) = \phi_r - \int_0^{z_*} \frac{z}{\sqrt{h(z)(z_*^2-z^2)}} = \phi_r - z_h\log\left(\frac{z_h + z_*}{\sqrt{z_h^2 - z_*^2}}\right).
\end{equation}

Thus, we may write the turnaround point in terms of $\phi_r$ and $z_h$ as,
\begin{equation}
z_* = z_h \tanh\left(\frac{\phi_r}{z_h}\right).\label{turnaround}
\end{equation}

Equipped with \eqref{turnaround}, the area of the exterior surface may be written as a function of $\phi_r$.
\begin{equation}
\tilde{A}_E = \int_0^{z_*} dz \frac{z_*}{z\sqrt{h(z)(z_*^2 - z^2)}}.
\end{equation}

At this point, note that the Hartman-Maldacena surface at $t_0 = 0$ consists \textit{solely} of two exterior pieces---the interior is simply a point. Consequently, there is no area contribution from behind the horizon. Furthermore, as discussed in \cite{Hartman:2013qma}, this contribution grows in time, thus implying a phase transition between the (overall) Hartman-Maldacena surface and exterior surface, so long as, at $t_0 = 0$,
\begin{equation}
\tilde{A}_{HM} < \tilde{A}_E.
\end{equation}

As done in \cite{Geng:2020qvw}, the fastest way to check that this occurs for sufficiently large $\phi_r$ is to make a numerical plot of the area difference (in which the UV divergences of the two areas cancel),
\begin{equation}
\Delta\tilde{A} = \tilde{A}_E - \tilde{A}_{HM} = \int_{0}^{z_*} dz \frac{z_* - \sqrt{z_*^2 - z^2}}{z\sqrt{h(z)(z_*^2 - z^2)}} - \int_{z_*}^{z_h} \frac{dz}{z\sqrt{h(z)}}.
\end{equation}

\begin{figure}
\centering
\includegraphics[scale=0.75]{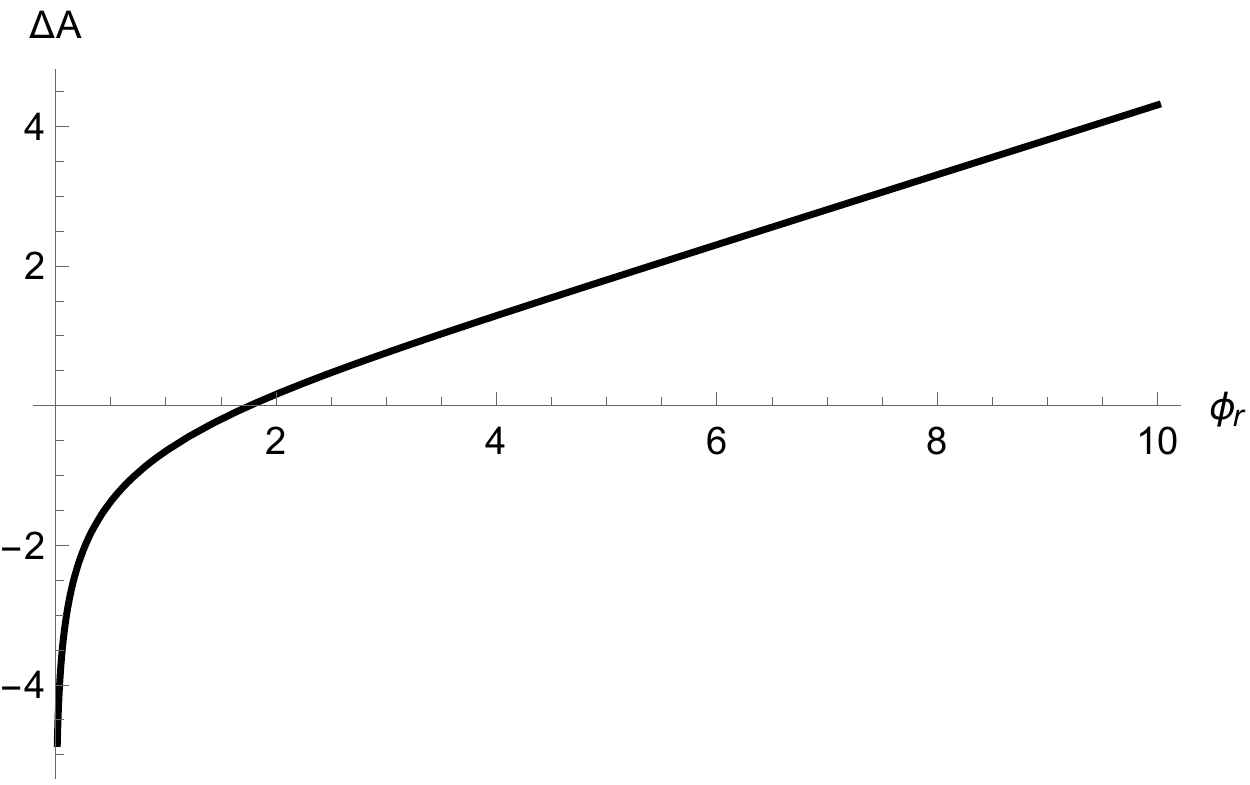}
\caption{A numerical plot of the area difference $\Delta\tilde{A}$ between the RT candidates shown in Figure \ref{figs:braneworldCFT}, as a function of the endpoint of $\mathcal{R}(t_0)$. We have fixed $z_h = 2$. For sufficiently large $\phi_r$, this plot confirms that the Hartman-Maldacena surface is minimal at $t_0 = 0$.}
\label{figs:numPlotCFT}
\end{figure}

We present this plot in Figure \ref{figs:numPlotCFT}. The Page curve is thus indeed the one depicted in Figure \ref{figs:eternalEntropy}. Additionally, this confirms the statement in \cite{Almheiri:2019yqk} that, for eternal AdS black holes in AdS/CFT, there are no islands at $t = 0$ but, after the phase transition, there are islands starting in the exterior regions and reaching behind the horizon.

\end{appendices}

\bibliographystyle{jhep}
\bibliography{multi}
\end{document}